# Extensions of smoothing via taut strings[*]

**Lutz Dümbgen**[†]

*University of Bern*
*Institute for Mathematical Statistics and Actuarial Science*
*Alpeneggstraße 22*
*CH-3012 Bern*
*e-mail:* lutz.duembgen@stat.unibe.ch

**Arne Kovac**[‡]

*University of Bristol*
*Department of Mathematics*
*University Walk*
*Bristol BS8 1TW, UK*
*e-mail:* a.kovac@bristol.ac.uk

**Abstract:** Suppose that we observe independent random pairs $(X_1, Y_1)$, $(X_2, Y_2)$, ..., $(X_n, Y_n)$. Our goal is to estimate regression functions such as the conditional mean or $\beta$–quantile of $Y$ given $X$, where $0 < \beta < 1$. In order to achieve this we minimize criteria such as, for instance,

$$\sum_{i=1}^{n} \rho(f(X_i) - Y_i) + \lambda \cdot \mathrm{TV}(f)$$

among all candidate functions $f$. Here $\rho$ is some convex function depending on the particular regression function we have in mind, $\mathrm{TV}(f)$ stands for the total variation of $f$, and $\lambda > 0$ is some tuning parameter. This framework is extended further to include binary or Poisson regression, and to include localized total variation penalties. The latter are needed to construct estimators adapting to inhomogeneous smoothness of $f$. For the general framework we develop noniterative algorithms for the solution of the minimization problems which are closely related to the taut string algorithm (cf. Davies and Kovac, 2001). Further we establish a connection between the present setting and monotone regression, extending previous work by Mammen and van de Geer (1997). The algorithmic considerations and numerical examples are complemented by two consistency results.

Received March 2008.

---

[*]The authors thank Axel Munk and an Associate Editor for constructive comments.
[†]Work supported by Swiss National Science Foundation
[‡]Work supported by Sonderforschungsbereich 475, University of Dortmund

## 1. Introduction

Suppose that we observe pairs $(x_1, Y_1), (x_2, Y_2), \ldots, (x_n, Y_n)$ with fixed numbers $x_1 \le x_2 \le \cdots \le x_n$ and independent random variables $Y_1, Y_2, \ldots, Y_n$. We assume that the distribution function of $Y_i$ depends on $x_i$, i.e.

$$\mathbb{P}(Y_i \le z) = F(z \,|\, x_i)$$

for some unknown family of distribution functions $F(\cdot \,|\, x)$, $x \in \mathbb{R}$. Often one is interested in certain features of these distribution functions. Examples are the mean function $\mu$ with

$$\mu(x) := \int y \, F(dy \,|\, x)$$

and, for some $\beta \in (0, 1)$, the $\beta$–quantile function $Q_\beta$, where $Q_\beta(x)$ is any number $z$ such that

$$F(z- \,|\, x) \le \beta \le F(z \,|\, x).$$

This paper treats estimation of such regression functions utilizing certain roughness penalties. The literature about penalized regression estimators is vast and still growing. As a good starting point we recommend Antoniadis and Fan (2001), van de Geer (2001), Huang (2003), and the references therein. A first possibility is to minimize a functional of the form

$$T(f) := \sum_{i=1}^{n} \rho(f(x_i) - Y_i) + \lambda \cdot \mathrm{TV}(f) \qquad (1)$$

over all functions $f$ on the real line. Here $\rho$ is some convex function measuring the size of the residual $Y_i - f(x_i)$ and depending on the particular feature we have in mind. Moreover, $\mathrm{TV}(f)$ denotes the total variation of $f$, that is the supremum of $\sum_{j=1}^{m-1} |f(z_{j+1}) - f(z_j)|$ over all integers $m > 1$ and numbers $z_1 < z_2 < \cdots < z_m$, while $\lambda > 0$ is some tuning parameter.

**Example I (means).** In order to estimate the mean function $\mu$, one can take

$$\rho(z) := z^2/2.$$

This particular case has been treated in detail by Mammen and van de Geer (1997) and Davies and Kovac (2001); see also the remark following Lemma 2.2. In particular, the latter authors describe an algorithm with running time $O(n)$, the taut string method, to minimize the functional $T$ above.

**Example II (quantiles).** For the estimation of a quantile function $Q_\beta$ one can take

$$\rho(z) = \rho_\beta(z) := |z|/2 - (\beta - 1/2)z = \begin{cases} (1-\beta)z & \text{if } z \ge 0, \\ \beta|z| & \text{if } z \le 0. \end{cases}$$

Of particular interest is the case $\beta = 1/2$. Then $\rho(z) = |z|/2$, and $Q_{1/2}$ is the conditional median function. This particular functional has also been suggested by Simpson, He and Liu in their discussion of Chu et al. (1998) but, as far as we know, not been considered in more detail later on. However, similar functionals using a discretisation of the total variation of the first derivative as a penalty have been studied by Koenker, Ng and Portnoy (1994) or in two dimensions by Koenker and Mizera (2004). They employ linear programming techniques like interior point methods to find solutions to the resulting minimisation problems.

A primary goal of the present paper is to extend the classical taut string algorithm to other situations such as Example II, or binary and Poisson regression. Compared to the linear programming techniques mentioned above, the generalized taut string method has the advantage of being computationally faster and more stable. In the specific case of Example II it is possible to calculate a solution in time $O(n \log(n))$. Note that the original algorithm yields piecewise constant functions. On each constant interval the function value is equal to the mean of the corresponding observations, except for local extrema of the fit. In their discussion of Davies and Kovac (2001), Mammen and van de Geer mention the possibility to replace sample means just by sample quantiles, in order to treat Example II. However, the present authors realized that the extension is not that straightforward.

The remainder of this paper is organized as follows: In Section 2 we describe an extension of the function $T$ above such that it covers also other models such as binary and Poisson regression. In addition we replace the penalty term $\lambda \cdot \mathrm{TV}(f)$ by a more flexible roughness measure which allows local adaptation to varying smoothness of the underlying regression function. Then we derive necessary and sufficient conditions for a function $\hat{f}$ to minimize our functional. In that context we also establish a connection to monotone regression which is useful for understanding adaptivity properties of the procedure. This generalizes findings of Mammen and van de Geer (1997) for the least squares case. In Sections 3 and 4 we derive generalized taut string algorithms, extending the algorithm described by Davies and Kovac (2001). While Section 3 covers continuously differentiable functions $\rho$, Section 4 is for general $\rho$ and, in particular, Example II. Section 5 explains how our tuning parameters, e.g. $\lambda$ in (1), may be chosen. Section 6 presents some numerical examples of our methods. In Section 7 we complement the algorithmic considerations with two consistency results which are of independent interest. One of them entails uniform consistency of monotone regression estimators, while the other applies to arbitrary estimators such that the corresponding residuals satisfy a certain multiscale criterion. Both results are a first step towards a detailed asymptotic analysis of taut string and related methods. All longer proofs are deferred to Section 8.

## 2. The general setting

For simplicity we assume throughout this paper that $x_1 < x_2 < \cdots < x_n$. In Section 3.3 we describe briefly possible modifications to deal with potential ties among the $x_i$.

### 2.1. The target functional

We often identify a function $f : \mathbb{R} \to \mathbb{R}$ with the vector $\boldsymbol{f} = (f_i)_{i=1}^n := (f(x_i))_{i=1}^n$. Our aim is to minimize functionals of the form

$$T(\boldsymbol{f}) = T_{\boldsymbol{\lambda}}(\boldsymbol{f}) := \sum_{i=1}^n R_i(f_i) + \sum_{j=1}^{n-1} \lambda_j |f_{j+1} - f_j|$$

over all vectors $\boldsymbol{f} \in \mathbb{R}^n$, where $\boldsymbol{\lambda} \in (0, \infty)^{n-1}$ is a given vector of tuning parameters while the $R_i$ are random functions depending on the data. In general we assume the following two conditions to be satisfied:

**(A.1)** For each index $i$, the function $R_i : \mathbb{R} \to \mathbb{R}$ is convex.
**(A.2)** $T$ is coercive, i.e. $T(\boldsymbol{f}) \to \infty$ as $\|\boldsymbol{f}\| \to \infty$.

Condition (A.1) entails that $T$ is a continuous and convex functional on $\mathbb{R}^n$, so the additional Condition (A.2) guarantees the existence of a minimizer $\hat{\boldsymbol{f}}$ of $T$. This will be our estimator for the regression function of interest, evaluated at the design points $x_i$.

The special functional $T$ in (1) corresponds to $\lambda_1 = \cdots = \lambda_{n-1} = \lambda$ and $R_i(z) := \rho(z - Y_i)$. Here our Conditions (A.1–2) are satisfied if $\rho(z) \to \infty$ as $|z| \to \infty$. Two additional examples for $R_i$ follow.

**Example III (Poisson regression).** Suppose that $Y_i \in \{0, 1, 2, \ldots\}$ has a Poisson distribution with mean $\mu(x_i) > 0$, and let

$$R_i(z) := \exp(z) - zY_i.$$

These functions are strictly convex with $R_i \geq Y_i \log(e/Y_i)$. Thus $T$ is even strictly convex, and elementary considerations reveal that it is coercive if $Y_i > 0$ for at least one index $i$. In that case we end up with a unique penalized maximum likelihood estimator $\hat{f}$ of $\log \mu$.

**Example IV (Binary regression).** Similarly let $Y_i \in \{0, 1\}$ with mean $\mu(x_i) \in (0, 1)$, and define

$$R_i(z) := -Y_i z + \log(1 + \exp(z)).$$

Here $R_i > 0$, and again $T$ is strictly convex. It is coercive if the $Y_i$ are not all identical, and the minimizer $\hat{\boldsymbol{f}}$ of $T$ may be viewed as a penalized maximum likelihood estimator of $\mathrm{logit}(\mu) := \log(\mu/(1-\mu))$.

### 2.2. Characterizations of the solution

As mentioned before, Conditions (A.1–2) guarantee the existence of a minimizer $\hat{\boldsymbol{f}}$ of $T$. In the present subsection we derive various characterizations of such minimizers, assuming only Condition (A.1).

By convexity of $T$, a vector $\hat{\boldsymbol{f}} \in \mathbb{R}^n$ minimizes $T$ if, and only if, all directional derivatives at $\hat{\boldsymbol{f}}$ are non-negative, i.e.

$$DT(\hat{\boldsymbol{f}}, \boldsymbol{\delta}) := \lim_{\varepsilon \downarrow 0} \frac{T(\hat{\boldsymbol{f}} + \varepsilon \boldsymbol{\delta}) - T(\hat{\boldsymbol{f}})}{\varepsilon} \geq 0 \quad \text{for any } \boldsymbol{\delta} \in \mathbb{R}^n. \qquad (2)$$

More specifically, let $R'_i(z\pm)$ be the left- and right-sided derivatives of $R_i$ at $z$, i.e.

$$R'_i(z\pm) := \lim_{\varepsilon \downarrow 0} \frac{R_i(z \pm \varepsilon) - R_i(z)}{\pm \varepsilon}.$$

Then $DT(\hat{\boldsymbol{f}}, \boldsymbol{\delta})$ equals

$$\sum_{i:\delta_i > 0} R'_i(\hat{f}_i+) \delta_i + \sum_{i:\delta_i < 0} R'_i(\hat{f}_i-) \delta_i$$
$$+ \sum_{j=1}^{n-1} \lambda_j \Big( \text{sign}(\hat{f}_{j+1} - \hat{f}_j)(\delta_{j+1} - \delta_j) + 1\{\hat{f}_{j+1} = \hat{f}_j\}|\delta_{j+1} - \delta_j| \Big),$$

where $1\{\ldots\}$ denotes the indicator function.

Plugging in various special vectors $\boldsymbol{\delta}$ reveals valuable information about minimizers $\hat{\boldsymbol{f}}$. In particular, for indices $1 \leq j \leq k \leq n$ let

$$\boldsymbol{\delta}^{(jk)} := \Big( 1\{j \leq i \leq k\} \Big)_{i=1}^n.$$

Then

$$DT(\hat{\boldsymbol{f}}, +\boldsymbol{\delta}^{(jk)}) = \sum_{i=j}^{k} R'_i(\hat{f}_i+) - \lambda_{j-1} \underline{\text{sign}}(\hat{f}_{j-1} - \hat{f}_j) - \lambda_k \underline{\text{sign}}(\hat{f}_{k+1} - \hat{f}_k),$$

$$DT(\hat{\boldsymbol{f}}, -\boldsymbol{\delta}^{(jk)}) = -\sum_{i=j}^{k} R'_i(\hat{f}_i-) + \lambda_{j-1} \overline{\text{sign}}(\hat{f}_{j-1} - \hat{f}_j) + \lambda_k \overline{\text{sign}}(\hat{f}_{k+1} - \hat{f}_k).$$

Here

$$\underline{\text{sign}}(z) := 1\{z > 0\} - 1\{z \leq 0\} \quad \text{and} \quad \overline{\text{sign}}(z) := 1\{z \geq 0\} - 1\{z < 0\},$$

and throughout this paper we set $v_0 := v_{m+1} := 0$ for any vector $\boldsymbol{v} = (v_i)_{i=1}^m \in \mathbb{R}^m$. In particular, $\lambda_0 := \lambda_n := 0$. Consequently, applying (2) to $\pm \boldsymbol{\delta}^{(jk)}$ yields the key inequalities

$$\begin{aligned}
\sum_{i=j}^{k} R'_i(\hat{f}_i+) &\geq \lambda_{j-1} \underline{\text{sign}}(\hat{f}_{j-1} - \hat{f}_j) + \lambda_k \underline{\text{sign}}(\hat{f}_{k+1} - \hat{f}_k) \quad \text{and} \\
\sum_{i=j}^{k} R'_i(\hat{f}_i-) &\leq \lambda_{j-1} \overline{\text{sign}}(\hat{f}_{j-1} - \hat{f}_j) + \lambda_k \overline{\text{sign}}(\hat{f}_{k+1} - \hat{f}_k).
\end{aligned} \qquad (3)$$

These considerations yield already one part of the following result.

**Lemma 2.1** *A vector $\hat{\boldsymbol{f}} \in \mathbb{R}^n$ minimizes $T$ if, and only if, (3) holds for all $1 \leq j \leq k \leq n$.*

In case of differentiable functions $R_i$ there is a simpler characterization of a minimizer of $T$:

**Lemma 2.2** *Suppose that all functions $R_i$ are differentiable on $\mathbb{R}$. Then a vector $\hat{\boldsymbol{f}} \in \mathbb{R}^n$ minimizes $T$ if, and only if, for $1 \le k \le n$,*

$$\sum_{i=1}^{k} R_i'(\hat{f}_i) \begin{cases} \in & [-\lambda_k, \lambda_k], \\ = & \lambda_k \quad \text{if } k < n \text{ and } \hat{f}_k < \hat{f}_{k+1}, \\ = & -\lambda_k \quad \text{if } k < n \text{ and } \hat{f}_k > \hat{f}_{k+1}. \end{cases} \quad (4)$$

For $k = n$, it follows from $\lambda_n = 0$ that Condition (4) amounts to

$$\sum_{i=1}^{n} R_i'(\hat{f}_i) = 0.$$

Note that in the classical case, $R_i(z) = (z - Y_i)^2/2$ and $\sum_{i=1}^{k} R_i'(\hat{f}_i) = \sum_{i=1}^{k} \hat{f}_i - \sum_{i=1}^{k} Y_i$. Thus our result entails Mammen and van de Geer's (1997) finding that the solution $\hat{\boldsymbol{f}}$ may be represented as the derivative of a taut string connecting the points $(0, 0)$ and $(n, \sum_{i=1}^{n} Y_i)$ and forced to lie within a tube centered at the points $(k, \sum_{i=1}^{k} Y_i)$, $1 \le k < n$.

In the general setting treated here, there are no longer taut strings, but the solutions can still be characterized by a tube. This is illustrated in the left panels of Figure 1 with a small example. The upper panel shows a data set of size $n = 25$ (with $x_i = i$) and the approximation $\hat{f}$ obtained from the functional $T$ in (1) with $\rho(z) := \sqrt{0.1^2 + z^2}$ and $\lambda = 2$. This function $\rho(r)$ may be viewed as a smoothed version of $|z|$ with

$$R_i'(z) = \frac{z - Y_i}{\sqrt{0.1^2 + (z - Y_i)^2}}$$

being similar to $\text{sign}(z - Y_i)$. The lower panel shows the cumulative sums of the "residuals" $R_i'(\hat{f}_i)$. As predicted by Lemma 2.2, these sums are always between $-\lambda$ and $\lambda$, and they touch these boundaries whenever the value of $\hat{f}$ changes.

### 2.3. Bounding the range of the solutions

Sometimes it is helpful to know a priori some bounds for any minimizer $\hat{\boldsymbol{f}}$ of $T$. We start with a rather obvious fact: Suppose that there are numbers $z_\ell < z_r$ such that for $i = 1, \ldots, n$,

$$R_i'(z+) < 0 \quad \text{if} \quad z < z_\ell,$$
$$R_i'(z-) > 0 \quad \text{if} \quad z > z_r.$$

Then any minimizer of $T$ belongs to $[z_\ell, z_r]^n$. For if $\boldsymbol{f} \in \mathbb{R}^n \setminus [z_\ell, z_r]^n$, one can easily verify that replacing $\boldsymbol{f}$ with $\big(\min(\max(f_i, z_\ell), z_r)\big)_{i=1}^{n}$ yields a strictly smaller value of $T(\boldsymbol{f})$. In case of differentiable functions $R_i$ an even stronger statement is true:

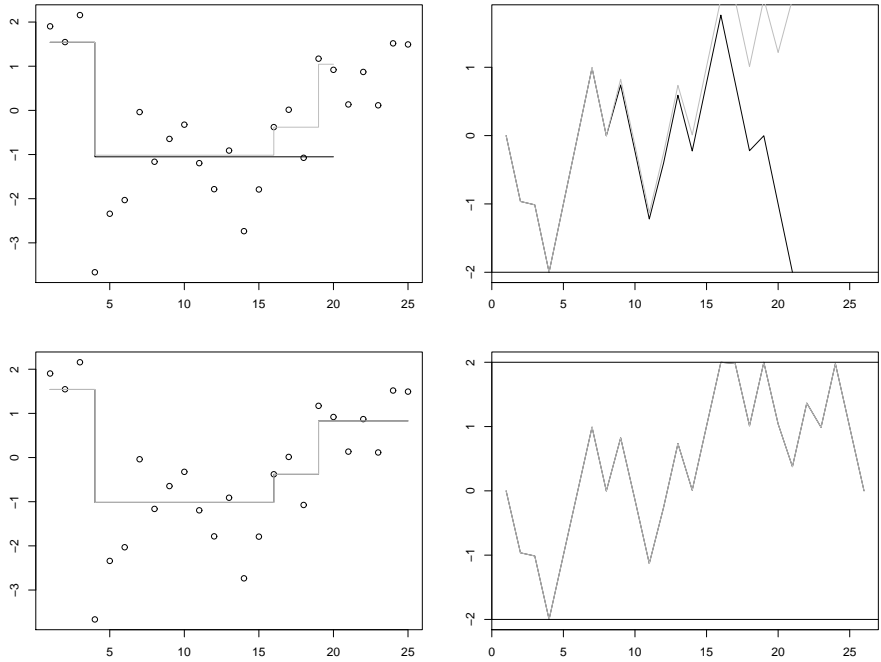

Fig 1. *Illustration of Lemma 2.2 and Conditions (C.$1_K$) and (C.$2_{f\&g,K}$) of Algorithm I.*

**Lemma 2.3** *Suppose that the functions $R_i$ are differentiable such that for certain numbers $z_\ell < z_r$,*

$$R_i'(z_\ell) \le 0 \quad \text{for all } i \quad \text{with at least one strict inequality,}$$
$$R_i'(z_r) \ge 0 \quad \text{for all } i \quad \text{with at least one strict inequality.}$$

*Then any minimizer of $T$ is contained in $(z_\ell, z_r)^n$.*

In the special case of $R_i(z) = \rho(z - Y_i)$ we reach the following conclusions: If 0 is the unique minimizer of $\rho$ over $\mathbb{R}$, then any minimizer of $T$ belongs to

$$\bigl[\min(Y_1, \ldots, Y_n), \max(Y_1, \ldots, Y_n)\bigr]^n.$$

If in addition $\rho$ is differentiable and $\boldsymbol{Y} = (Y_i)_{i=1}^n$ is non-constant, then any minimizer of $T$ belongs even to

$$\bigl(\min(Y_1, \ldots, Y_n), \max(Y_1, \ldots, Y_n)\bigr)^n.$$

### *2.4. A link to monotone regression*

An interesting alternative to smoothness assumptions and roughness penalties is to assume monotonicity of the underlying regression function $f$ on certain intervals. For instance, if $f$ is assumed to be isotonic, one could determine an estimator $\check{\boldsymbol{f}}$ minimizing $\sum_{i=1}^n R_i(\check{f}_i)$ under the constraint that $\check{f}_1 \leq \check{f}_2 \leq \cdots \leq \check{f}_n$. The next theorem shows that our penalized estimators $\hat{\boldsymbol{f}}$ often coincide locally with monotone estimators.

**Theorem 2.4** *Suppose that $1 < a \leq b < n$ such that*

$$\lambda_{a-1} = \lambda_a = \cdots = \lambda_b \quad \text{and} \quad \hat{f}_{a-1} < \hat{f}_a \leq \cdots \leq \hat{f}_b < \hat{f}_{b+1}.$$

*Then $(\hat{f}_i)_{i=a}^b$ minimizes $\sum_{i=a}^b R_i(f_i)$ among all vectors $(f_i)_{i=a}^b$ satisfying $f_a \leq \cdots \leq f_b$. An analogous statement holds for antitonic fits.*

## 3. Computation in case of regular functions $R_i$

### 3.1. A general algorithm for strongly coercive functions $R_i$

In this subsection we present an algorithm for the minimization of the functional $T$ in case of (A.1) together with the additional constraint that

**(A.3)** all functions $R_i$ are continuously differentiable with

$$\lim_{z \to -\infty} R'_i(z) = -\infty \quad \text{and} \quad \lim_{z \to \infty} R'_i(z) = \infty.$$

Obviously the latter condition is satisfied in Example I. Note that Conditions (A.1) and (A.3) imply Condition (A.2). Moreover, $R'_i : \mathbb{R} \to \mathbb{R}$ is isotonic (i.e. non-decreasing), continuous and surjective.

**The algorithm's principle.** The idea of our algorithm is to compute inductively for $K = 1, 2, \ldots, n$ a vector $(\hat{f}_i)_{i=1}^K$ such that Condition (4) holds for $1 \leq k \leq K$, where $\hat{f}_{K+1}$ may be defined arbitrarily.

Precisely, inductively for $K = 1, 2, \ldots, n$ we compute *two* candidate vectors $\boldsymbol{f} = (f_i)_{i=1}^K$ and $\boldsymbol{g} = (g_i)_{i=1}^K$ in $\mathbb{R}^K$ such that at the end of step $K$ the following three conditions are satisfied:

**(C.1$_K$)** There exists an index $k_o = k_o(\boldsymbol{f}, \boldsymbol{g}) \in \{0, 1, \ldots, K\}$ such that

$$g_i - f_i \begin{cases} = 0 & \text{for } 1 \leq i \leq k_o, \\ > 0 & \text{for } k_o < i \leq K. \end{cases}$$

Moreover, $f_i$ is antitonic (i.e. non-increasing) and $g_i$ is isotonic in $i \in \{k_o + 1, \ldots, K\}$.

**(C.2$_{g,K}$)** For $1 \leq k \leq K$,

$$\sum_{i=1}^k R'_i(g_i) \leq \lambda_k \quad \text{with equality if} \quad \begin{cases} k < K \text{ and } g_k < g_{k+1}, \\ k = K. \end{cases}$$

**(C.2$_{f,K}$)** For $1 \leq k \leq K$,

$$\sum_{i=1}^k R'_i(f_i) \geq -\lambda_k \quad \text{with equality if} \quad \begin{cases} k < K \text{ and } f_k > f_{k+1}, \\ k = K. \end{cases}$$

Note that Conditions (C.1$_K$) and (C.2$_{f\&g,K}$) imply the following fact:

**(C.3$_K$)**    Let $\Lambda_o = \Lambda_o(\boldsymbol{f},\boldsymbol{g}) := \sum_{i=1}^{k_o} R'_i(f_i)$. If $1 \leq k_o < K$, then either

$$f_{k_o} = g_{k_o} \geq g_{k_o+1} > f_{k_o+1} \quad \text{and} \quad \Lambda_o = -\lambda_{k_o},$$

or

$$g_{k_o} = f_{k_o} \leq f_{k_o+1} < g_{k_o+1} \quad \text{and} \quad \Lambda_o = \lambda_{k_o}.$$

When the algorithm finishes with $K = n$, a solution $\hat{\boldsymbol{f}} \in \mathbb{R}^n$ may be obtained as follows: If $k_o = n$, $\hat{\boldsymbol{f}} := \boldsymbol{f} = \boldsymbol{g}$ satisfies the conditions of Lemma 2.2. If $k_o < n$, we define

$$\hat{f}_i := \begin{cases} f_i = g_i & \text{for } 1 \leq i \leq k_o \\ r & \text{for } k_o < i \leq n \end{cases}$$

with some number $r \in [f_{k_o+1}, g_{k_o+1}]$. This definition entails that $f_i \leq \hat{f}_i \leq g_i$ for all indices $i$, while $\hat{f}_i$ is constant in $i > k_o$. Hence one can easily deduce from Conditions (C.2$_{f\&g,n}$) that $\hat{\boldsymbol{f}}$ satisfies the conditions of Lemma 2.2. To ensure a certain optimality property described later, in case of $1 \leq k_o < n$ we choose

$$r := \begin{cases} g_{k_o+1} & \text{if } f_{k_o} = g_{k_o} \geq g_{k_o+1}, \\ f_{k_o+1} & \text{if } g_{k_o} = f_{k_o} \leq f_{k_o+1}. \end{cases}$$

Conditions (C.1$_K$) and (C.2$_{f\&g,K}$) are illustrated in the right panels of Figure 1 which show the same example as the left panels that were discussed in the previous Section. Strictly speaking, Condition (A.3) is not satisfied here, but one can enforce it by adding $\min(z - c_1, 0)^2 + \max(z - c_2, 0)^2$ to $R_i(z) = \rho(z - Y_i)$ with arbitrary constants $c_1 \leq \min(Y_1, \ldots, Y_n)$ and $c_2 \geq \max(Y_1, \ldots, Y_n)$. This modification does not alter the solution which is contained in $[c_1, c_2]^n$; see Section 2.3.

The solid and grey lines in the upper panel represent $\boldsymbol{f}$ and $\boldsymbol{g}$, respectively, where $K = 20$ and $k_o = 3$. The lower panel shows the corresponding cumulative sums of the numbers $R'_i(f_i)$ and $R'_i(g_i)$, $1 \leq i \leq K$.

**Some auxiliary functions and terminology.**    Later on we shall work with partitions of the set $\{1, 2, \ldots, n\}$ into index intervals and functions (vectors) which are constant on these intervals. To define the latter vectors efficiently we define

$$R'_{jk} := \sum_{i=j}^{k} R'_i$$

for indices $1 \leq j \leq k \leq n$. Again, $R'_{jk}$ is continuous and isotonic on $\mathbb{R}$ with limits $R'_{jk}(\pm\infty) = \pm\infty$. Further, for real numbers $t$ let

$$\underline{M}_{jk}(t) := \min\{z \in \mathbb{R} : R'_{jk}(z) \geq t\},$$
$$\overline{M}_{jk}(t) := \max\{z \in \mathbb{R} : R'_{jk}(z) \leq t\}.$$

These quantities $\underline{M}_{jk}(t)$ and $\overline{M}_{jk}(t)$ are isotonic in $t$, where $R'_{jk}(z) = t$ for any real number $z \in [\underline{M}_{jk}(t), \overline{M}_{jk}(t)]$.

The following lemma summarizes basic properties of the auxiliary functions $\overline{M}_{jk}$ which are essential for Algorithm I below. The functions $\underline{M}_{jk}$ satisfy analogous properties.

**Lemma 3.1** *Let $1 \le j \le k < \ell \le m \le n$ be indices with $\ell = k + 1$. Further let $c, u, v \in \mathbb{R}$ such that*

$$R'_{jk}(c) = u.$$

**(a)** *If $c \ge \overline{M}_{\ell m}(v)$, then*

$$c \ge \overline{M}_{jm}(u+v) \ge \overline{M}_{\ell m}(v).$$

**(b)** *If $c > \overline{M}_{jm}(u+v)$, then*

$$\overline{M}_{jm}(u+v) \ge \overline{M}_{\ell m}(v).$$

For some vector $\boldsymbol{v} = (v_i)_{i=a}^b$, a maximal index interval $J \subset \{a, \ldots, b\}$ such that $v_i$ is constant in $i \in J$ will be called a "segment of $\boldsymbol{v}$".

**Algorithm I: Step 1.** For $K = 1$ we define

$$f_1 := \underline{M}_{1,1}(-\lambda_1) \quad \text{and} \quad g_1 := \overline{M}_{1,1}(\lambda_1).$$

Conditions (C.$1_1$) and (C.$2_{f\&g,1}$) are certainly satisfied.

**Algorithm I: Step $K + 1$.** Suppose that Conditions (C.$1_K$) and (C.$2_{f\&g,K}$) are satisfied for some $K \in \{1, \ldots, n-1\}$. Since $\lambda_K > 0$, one can easily derive from (C.$2_{f\&g,K}$) that $k_o < K$. Subsequently we construct new candidates $\tilde{\boldsymbol{f}} = (\tilde{f}_i)_{i=1}^{K+1}$ for $\boldsymbol{f}$ and $\tilde{\boldsymbol{g}} = (\tilde{g}_i)_{i=1}^{K+1}$ for $\boldsymbol{g}$. In this context, (C.$1_{K+1}$) and (C.$2_{\ldots,K+1}$) always denote conditions on the new vectors $\tilde{\boldsymbol{f}}$ and $\tilde{\boldsymbol{g}}$ in place of $\boldsymbol{f}$ and $\boldsymbol{g}$, respectively, while (C.$1_K$) and (C.$2_{\ldots,K}$) refer to the original $\boldsymbol{f}$ and $\boldsymbol{g}$.

**Initializing $\tilde{\boldsymbol{g}}$.** We set

$$\tilde{g}_i := \begin{cases} g_i & \text{for } i \le K, \\ \overline{M}_{K+1,K+1}(\lambda_{K+1} - \lambda_K) & \text{for } i = K + 1. \end{cases}$$

Since $\tilde{g}_i = g_i$ for all $i \le K$, one can easily verify that the inequality part of (C.$2_{g,K+1}$) is satisfied, and also $\sum_{i=1}^{K+1} R'_i(\tilde{g}_i) = \lambda_{K+1}$.

**Modifying $\tilde{\boldsymbol{g}}$.** Suppose that $\tilde{g}_i$ is not isotonic in $i > k_o$. By the previous construction of $\tilde{\boldsymbol{g}}$, this means that the two rightmost segments $\{j, \ldots, k\}$ and $\{\ell, \ldots, K+1\}$ of $(\tilde{g}_i)_{i=k_o+1}^{K+1}$ satisfy $\tilde{g}_k > \tilde{g}_\ell$, where $k_o < j \le k = \ell - 1 \le K$. In that case we replace $\tilde{g}_i$, $i \in \{j, \ldots, K+1\}$, with

$$\begin{cases} \overline{M}_{j,K+1}(\lambda_{K+1} - \lambda_{j-1}) & \text{if } j > k_o + 1, \\ \overline{M}_{j,K+1}(\lambda_{K+1} - \Lambda_o) & \text{if } j = k_o + 1. \end{cases}$$

This step is repeated, if necessary, until $\tilde{g}_i$ is isotonic in $i > k_o$. One can easily deduce from Part (a) of Lemma 3.1 below that throughout $\tilde{g}_i \le g_i$ for $k_o < i \le K$ while $\sum_{i=1}^{K+1} R'_i(\tilde{g}_i) = \lambda_{K+1}$. Hence the inequality part of Condition (C.2$_{g,K+1}$) continues to hold. The equality statements of Condition (C.2$_{g,K+1}$) are true as well, the only possible exception being $k = k_o$. This exceptional case may occur only if $\tilde{g}_{k_o+1} < g_{k_o+1}$, and by our construction of $\tilde{g}$, this entails $\tilde{g}_i$ being constant in $i \in \{k_o + 1, \ldots, K+1\}$.

**Initializing $\tilde{f}$.** We set

$$\tilde{f}_i := \begin{cases} f_i & \text{for } i \le K, \\ \underline{M}_{K+1,K+1}(-\lambda_{K+1} + \lambda_K) & \text{for } i = K+1. \end{cases}$$

Again the inequality part of (C.2$_{f,K+1}$) is satisfied, and also $\sum_{i=1}^{K+1} R'_i(\tilde{f}_i) = -\lambda_{K+1}$.

**Modifying $\tilde{f}$.** Suppose that $\tilde{f}_i$ is not antitonic in $i > k_o$. This means that the two rightmost segments $\{j, \ldots, k\}$ and $\{\ell, \ldots, K+1\}$ of $(\tilde{f}_i)_{i=k_o+1}^{K+1}$ satisfy $\tilde{f}_k < \tilde{f}_\ell$, where $k_o < j \le k = \ell - 1 \le K$. In that case we replace $\tilde{f}_i$, $i \in \{j, \ldots, K+1\}$, with

$$\begin{cases} \underline{M}_{j,K+1}(-\lambda_{K+1} + \lambda_{j-1}) & \text{if } j > k_o + 1, \\ \underline{M}_{j,K+1}(-\lambda_{K+1} - \Lambda_o) & \text{if } j = k_o + 1. \end{cases}$$

This step is repeated, if necessary, until $\tilde{f}_i$ is antitonic in $i > k_o$. Throughout, $\tilde{f}_i \ge f_i$ for $k_o < i \le K$ while $\sum_{i=1}^{K+1} R'_i(\tilde{f}_i) = -\lambda_{K+1}$. Hence the inequality part of Condition (C.2$_{f,K+1}$) continues to be satisfied. The equality statements of Condition (C.2$_{f,K+1}$) are true as well, the only possible exception being $k = k_o$. This exceptional case may occur only if $\tilde{f}_{k_o+1} > f_{k_o+1}$, and this entails that $\tilde{f}_i$ is constant in $i > k_o$.

This step is illustrated in Figure 2 which shows again the example from Figure 1. Here $K = 17$ and the panels show from left to right the initialisation of $\tilde{f}$, the first modification of $\tilde{f}$ and the second and final modification. The panels in the upper row show the data and the approximations while the panels in the lower row show the corresponding cumulative sums of the numbers $R'_i(f_i)$ and $R'_i(g_i)$, $1 \le i \le K$.

**Final modification of $\tilde{f}$ and $\tilde{g}$.** Having completed the previous construction, we end up with vectors $\tilde{f}$ and $\tilde{g}$ satisfying the inequality parts of Conditions (C.2$_{f\&g,K+1}$). The equality parts are satisfied as well, with possible exceptions only for $k = k_o(\boldsymbol{f}, \boldsymbol{g})$. Moreover, $\tilde{f}_i$ is antitonic and $\tilde{g}_i$ is isotonic in $i > k_o$. Finally, our explicit construction entails that

$$f_{k_o+1} = \tilde{f}_{k_o+1} \quad \text{or} \quad f_{k_o+1} < \tilde{f}_{k_o+1} = \cdots = \tilde{f}_{K+1},$$

and

$$g_{k_o+1} = \tilde{g}_{k_o+1} \quad \text{or} \quad g_{k_o+1} > \tilde{g}_{k_o+1} = \cdots = \tilde{g}_{K+1}.$$

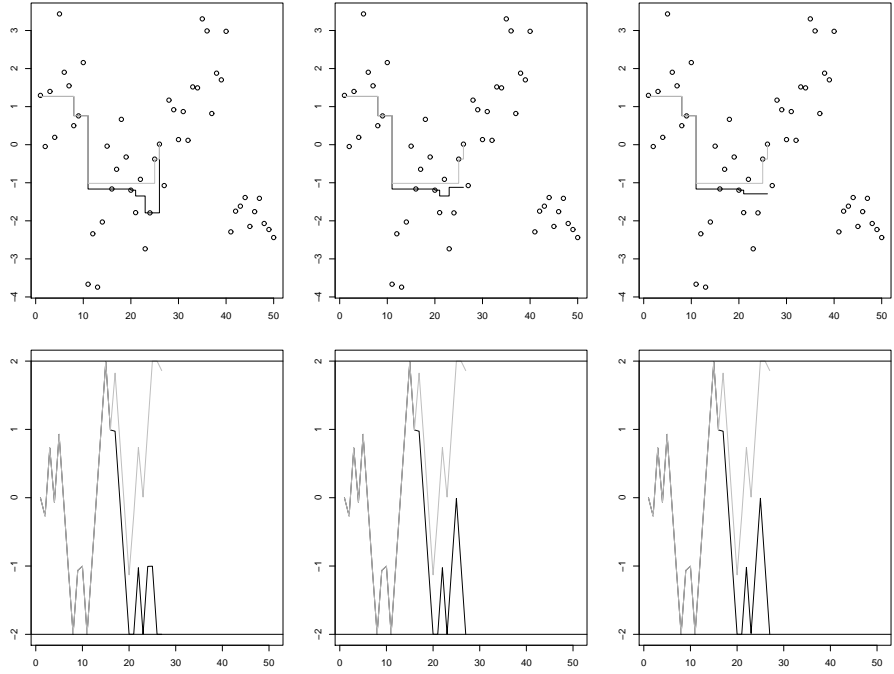

FIG 2. *Illustration of the modification step for $\tilde{\boldsymbol{f}}$ in Algorithm I.*

Suppose first that $\tilde{f}_{k_o+1} \leq \tilde{g}_{k_o+1}$. Then one can easily deduce from $(C.3_K)$ and the properties of $\tilde{\boldsymbol{f}}$, $\tilde{\boldsymbol{g}}$ just mentioned that Conditions $(C.1_{K+1})$ and $(C.2_{f\&g,K+1})$ are satisfied.

One particular instance of the previous situation is that both $\tilde{f}_i$ and $\tilde{g}_i$ are constant in $i > k_o$. For then

$$-\lambda_{K+1} \;=\; \sum_{i=1}^{K+1} R'_i(\tilde{f}_i) \;=\; \Lambda_o + R'_{k_o+1,K+1}(\tilde{f}_{k_o+1})$$

and

$$\lambda_{K+1} \;=\; \sum_{i=1}^{K+1} R'_i(\tilde{g}_i) \;=\; \Lambda_o + R'_{k_o+1,K+1}(\tilde{g}_{k_o+1}).$$

Now suppose that $\tilde{f}_{k_o+1} > \tilde{g}_{k_o+1}$. The previous considerations and our construction of $\tilde{\boldsymbol{f}}$ and $\tilde{\boldsymbol{g}}$ show that either

$$\tilde{f}_{K+1} \;<\; \tilde{f}_{k_o+1} = f_{k_o+1} \;>\; \tilde{g}_{k_o+1},$$

or

$$\tilde{g}_{K+1} \;>\; \tilde{g}_{k_o+1} = g_{k_o+1} \;<\; \tilde{f}_{k_o+1}.$$

We discuss only the former case, the latter case being handled analogously. Here Condition (C.$2_{f,K+1}$) is satisfied already. Let $\{k_o + 1, \ldots, k_1\}$ be the leftmost segment of $(\tilde{f}_i)_{i=k_o+1}^{K+1}$. Then we redefine $\tilde{g}$ as follows:

$$\tilde{g}_i := \begin{cases} f_i = g_i & \text{for } i \leq k_o, \\ \tilde{f}_{k_o+1} & \text{for } k_o < i \leq k_1, \\ \overline{M}_{k_1+1,K+1}(\lambda_{K+1} + \lambda_{k_1}) & \text{for } k_1 < i \leq K+1. \end{cases}$$

By assumption, $R'_{k_o+1,k_1}(\tilde{f}_{k_o+1}) = -\lambda_{k_1} - \Lambda_o$ and $\overline{M}_{k_o+1,K+1}(\lambda_{K+1} - \Lambda_o) < \tilde{f}_{k_o+1}$. Hence Part (b) of Lemma 3.1 entails that the new value of $\tilde{g}_{k_1+1}$ is not greater than $\overline{M}_{k_o+1,K+1}(\lambda_{K+1} - \Lambda_o)$, which is the old value of $\tilde{g}_{k_o+1} = \cdots = \tilde{g}_{K+1}$. Since $\tilde{f}_{k_o+1} = f_{k_o+1} < g_{k_o+1}$, we may conclude that the new vector $\tilde{g}$ still satisfies $\tilde{g}_i < g_i$ for $k_o < i \leq K$. Now one easily verifies that the new vector $\tilde{g}$ satisfies Condition (C.$2_{g,K+1}$).

It may happen that Condition (C.$1_{K+1}$) is still violated, i.e. $\tilde{g}_{k_1+1} < \tilde{f}_{k_1+1}$. In that case we repeat the previous update of $\tilde{g}$ with $k_1$ in place of $k_o$ and iterate this procedure until Condition (C.$1_{K+1}$) is satisfied as well.

This step is illustrated in Figure 3 which shows once more the example from Figure 1 and 2. Here $K = 23$ and the left panels show $\tilde{f}$ before the final modification while the result of the modification is shown in the right panel. As always the panels in the upper row show the data and the approximations while the panels in the lower row show the corresponding cumulative sums of the numbers $R'_i(f_i)$ and $R'_i(g_i)$, $1 \le i \le K$.

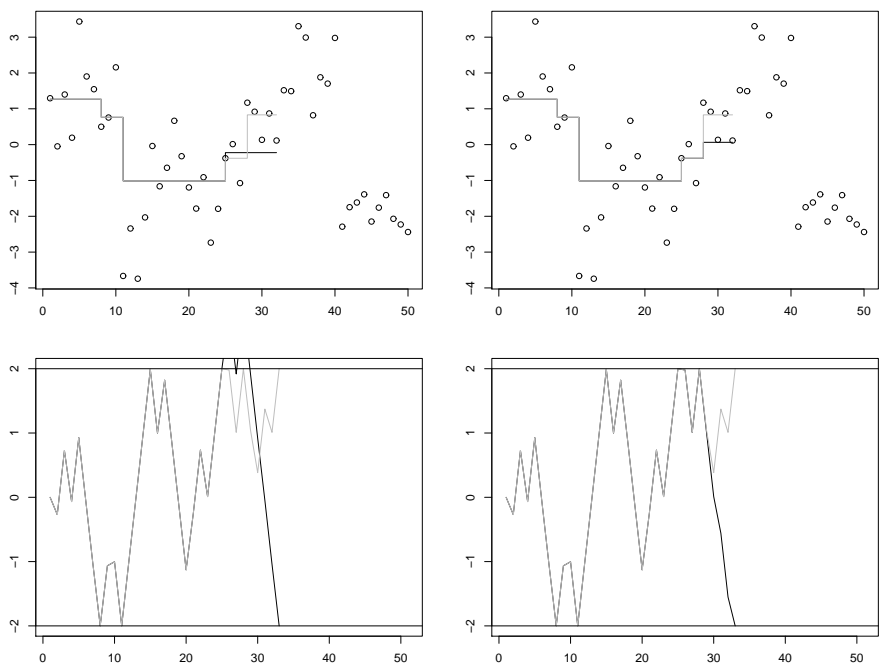

Fig 3. *Illustration of the final modification step for $\tilde{f}$ in Algorithm I.*

**An optimality property of Algorithm I.** The solution $\hat{\boldsymbol{f}}$ produced by Algorithm I is as simple as possible in a certain sense. For a vector $\boldsymbol{f} \in \mathbb{R}^n$, an index interval $\{j, \ldots, k\} \subset \{1, 2, \ldots, n\}$ with $j > 1$ or $k < n$ is called a

$$\begin{cases} \text{local maximum of } \boldsymbol{f} & \text{if } f_j = \cdots = f_k \ > \ \max_{i \in \mathcal{N}} f_i, \\ \text{local minimum of } \boldsymbol{f} & \text{if } f_j = \cdots = f_k \ < \ \min_{i \in \mathcal{N}} f_i, \end{cases}$$

where $\mathcal{N} := \{j-1, k+1\} \cap \{1, 2, \ldots, n\}$.

**Theorem 3.2** *Let $\hat{\boldsymbol{f}}$ be the vector produced by Algorithm I, and let $\boldsymbol{f}$ be any vector in $\mathbb{R}^n$ such that*

$$\left| \sum_{i=1}^{k} R'_i(f_i) \right| \ \le \ \lambda_k \quad \text{for } 1 \le k \le n. \tag{5}$$

*Then*

$$\max_{i \in \mathcal{J}} f_i \ \ge \ \max_{i \in \mathcal{J}} \hat{f}_i \quad \text{for any local maximum } \mathcal{J} \text{ of } \hat{\boldsymbol{f}},$$

$$\min_{i \in \mathcal{J}} f_i \ \le \ \min_{i \in \mathcal{J}} \hat{f}_i \quad \text{for any local minimum } \mathcal{J} \text{ of } \hat{\boldsymbol{f}}.$$

In particular the theorem shows that every vector $\boldsymbol{f}$ satisfying the tube condition (5) must have at least the same number of local extreme values as $\hat{\boldsymbol{f}}$.

### 3.2. Exponential families

Examples III and IV may be generalized as follows: Suppose that $Y_j$ has distribution $P_{f(x_j)}$ for some unknown real parameter $f(x_j)$, where $(P_\theta)_{\theta \in \mathbb{R}}$ is an exponential family with

$$\frac{dP_\theta}{d\nu}(y) \ = \ \exp\bigl(\theta y - b(\theta)\bigr)$$

for some measure $\nu$ on the real line.

In case of Poisson regression, $\nu$ is a discrete measure on $\{0, 1, 2, \ldots\}$ with weights $\nu(\{k\}) = 1/k!$, so that $b(\theta) = b'(\theta) = b''(\theta) = e^\theta$. For binomial regression we choose $\nu$ to be counting measure on $\{0, 1\}$, whence $b(\theta) = \log(1 + e^\theta)$, $b'(\theta) = e^\theta/(1 + e^\theta)$ and $b''(\theta) = b'(\theta)(1 - b'(\theta))$.

Note that in general, $b(\cdot)$ is infinitely often differentiable with $b'(\theta)$ and $b''(\theta)$ being the mean and variance, respectively, of the distribution $P_\theta$. We assume that $b''(\theta)$ is strictly positive for all $\theta \in \mathbb{R}$. Note also that $\nu(\mathbb{R} \setminus [y_{\min}, y_{\max}]) = 0$, where $y_{\min}$ and $y_{\max}$ are the infimum and supremum, respectively, of the set $\{b'(\theta) : \theta \in \mathbb{R}\}$.

Now we consider minimization of minus the log-likelihood function plus the localized total variation penalty, i.e.

$$R_i(z) := b(z) - zY_i.$$

By assumption, these functions are infinitely often differentiable with first derivative $R_i'(z) = b'(z) - Y_i$ and strictly positive second derivative $R_i''(z) = b''(z)$. Hence they satisfy Condition (A.1). Unfortunately they may fail to be coercive individually, and Condition (A.3) is not satisfied in general. However, one can show that Condition (A.2) is satisfied as soon as $\boldsymbol{Y}$ is non-trivial in the sense that

$$\min(Y_1, \ldots, Y_n) < \max(Y_1, \ldots, Y_n). \tag{6}$$

We will not pursue this here, because there is a simple solution of our estimation problem, at least in case of (6):

At first we apply the usual taut string method to the observations $Y_i$, i.e. we replace $R_i(z)$ with $\rho(z - Y_i)$, where $\rho(z) := z^2/2$. Let $\hat{\boldsymbol{f}}^{\text{LS}}$ be the resulting least squares fit. It follows from Lemma 2.3 that

$$\{\hat{f}_1^{\text{LS}}, \ldots, \hat{f}_n^{\text{LS}}\} \subset \big(\min(Y_1, \ldots, Y_n), \max(Y_1, \ldots, Y_n)\big).$$

Thus the vector $\hat{\boldsymbol{f}}$ with components

$$\hat{f}_i := (b')^{-1}(\hat{f}_i^{\text{LS}})$$

is well-defined and satisfies $\sum_{i=1}^k R_i'(\hat{f}_i) = \sum_{i=1}^k \rho'(\hat{f}_i^{\text{LS}} - Y_i)$. Hence applying Lemma 2.2 to $\hat{\boldsymbol{f}}^{\text{LS}}$ in the least squares setting entails the analogous conditions for $\hat{\boldsymbol{f}}$ in the maximum likelihood setting. This shows that $\hat{\boldsymbol{f}}$ is indeed the unique minimizer of $T$.

### 3.3. Ties among the $x_i$

For simplicity we assumed that $x_1 < x_2 < \cdots < x_n$. Now we relax this assumption temporarily to $x_1 \leq x_2 \leq \cdots \leq x_n$ and describe a possible modification of Algorithm I.

Let $x_{(1)} < x_{(2)} < \cdots < x_{(m)}$ be the distinct elements of $\{x_1, x_2, \ldots, x_n\}$, and let $i(k) := \max\{i : x_i = x_{(k)}\}$. Then we restrict our attention to vectors $\boldsymbol{f} \in \mathbb{R}^n$ such that $f_i = f_j$ for $i(k-1) < i < j \leq i(k)$, $1 \leq k \leq m$, where $i(0) := 0$. The target functional has to be rewritten as

$$T(\boldsymbol{f}) = \sum_{i=1}^n R_i(f_i) + \sum_{k=1}^{m-1} \lambda_j |f_{i(k+1)} - f_{i(k)}|.$$

Now Algorithm I uses induction on $K = 1, 2, \ldots, m$. In Step 1 we define

$$f_i := \underline{M}_{1,i(1)}(-\lambda_1) \quad \text{and} \quad g_i := \overline{M}_{1,i(1)}(\lambda_1) \quad \text{for } 1 \leq i \leq i(1).$$

In Step $K+1$ we aim for vectors $\tilde{\boldsymbol{f}}$ and $\tilde{\boldsymbol{g}}$ in $\mathbb{R}^{i(K+1)}$. The initial versions are given by

$$\tilde{g}_i := \begin{cases} g_i & \text{for } i \leq i(K), \\ \overline{M}_{i(K)+1,i(K+1)}(+\lambda_{K+1} - \lambda_K) & \text{for } i(K) < i \leq i(K+1), \end{cases}$$

$$\tilde{f}_i := \begin{cases} f_i & \text{for } i \leq i(K), \\ \underline{M}_{i(K)+1,i(K+1)}(-\lambda_{K+1} + \lambda_K) & \text{for } i(K) < i \leq i(K+1), \end{cases}$$

while the remainder of Step $K+1$ remains unchanged.

## 4. The case of arbitrary functions $R_i$

In this Section we describe how to calculate solutions to the general setting described in Section 2. In particular we investigate how to solve the quantile regression problem presented in Example II. Throughout this section we assume that Conditions (A.1–2) are satisfied, while some of the functions $R_i$ may fail to satisfy the regularity condition (A.3).

### 4.1. Approximating the $R_i$

We start with a general observation: For $\varepsilon > 0$ and $i \in \{1, 2, \ldots,\}$ let $R_{i,\varepsilon} : \mathbb{R} \to \mathbb{R}$ be a (data-driven) convex function such that

$$\lim_{\varepsilon \downarrow 0} R_{i,\varepsilon}(z) = R_i(z) \quad \text{for any } z \in \mathbb{R}. \tag{7}$$

The corresponding approximation $T_\varepsilon(\boldsymbol{f}) = T_{\boldsymbol{\lambda},\varepsilon}(\boldsymbol{f})$ to $T(\boldsymbol{f})$ equals $\sum_{i=1}^n R_{i,\varepsilon}(f_i) + \sum_{j=1}^{n-1} \lambda_j |f_{j+1} - f_j|$ and has the following properties:

**Theorem 4.1** *For sufficiently small $\varepsilon > 0$, the set $\hat{\mathcal{F}}_\varepsilon := \arg\min_{\boldsymbol{f} \in \mathbb{R}^n} T_\varepsilon(\boldsymbol{f})$ is nonvoid and compact. Moreover, it approximates the set $\hat{\mathcal{F}} := \arg\min_{\boldsymbol{f} \in \mathbb{R}^n} T(\boldsymbol{f})$ in the sense that*

$$\max_{\boldsymbol{f}_\varepsilon \in \hat{\mathcal{F}}_\varepsilon} \min_{\boldsymbol{f} \in \hat{\mathcal{F}}} \|\boldsymbol{f}_\varepsilon - \boldsymbol{f}\| \to 0 \quad as\ \varepsilon \downarrow 0.$$

If we find approximations $R_{i,\varepsilon}$ satisfying additionally Condition (A.3), we can minimize the target functional $T(\cdot)$ at least approximately by means of Algorithm I. One possible definition of such functions $R_{i,\varepsilon}$ is given by

$$R_{i,\varepsilon}(z) := \frac{1}{2\varepsilon} \int_{z-\varepsilon}^{z+\varepsilon} R_i(t)\, dt + \frac{\max(z - 1/\varepsilon, 0)^2}{2} + \frac{\min(z + 1/\varepsilon, 0)^2}{2}.$$

Here one can easily verify (7) and Condition (A.3) with $R_{i,\varepsilon}$ in place of $R_i$.

### 4.2. A non-iterative solution for Example II

Apparently the preceding considerations lead to an iterative procedure for minimizing $T$. But such a detour is not always necessary. In this section we derive an explicit combinatorial algorithm for Example II. Recall that for given $\boldsymbol{Y} = (Y_i)_{i=1}^n$, our goal is to minimize $T(\boldsymbol{f})$ with $R_i(z) := \rho_\beta(z - Y_i)$. Note first that
$$R_i'(z+) \;=\; 1\{Y_i \le z\} - \beta \quad \text{and} \quad R_i'(z-) \;=\; 1\{Y_i < z\} - \beta.$$
This indicates already that for quantile regression mainly the ranks of the vector $\boldsymbol{Y}$ matter. Precisely, we shall work with a permutation $\boldsymbol{Z} = (Z_i)_{i=1}^n$ of $(i)_{i=1}^n$ such that
$$\#\{i : Y_i < Y_j\} + 1 \;\le\; Z_j \;\le\; \#\{i : Y_i \le Y_j\} \quad \text{for } 1 \le j \le n. \tag{8}$$
That means, $\boldsymbol{Z}$ is a rank vector of $\boldsymbol{Y}$ but without the usual modification in case of ties. The usefulness of this will become clear later. Solving the original problem with $\boldsymbol{Z}$ in place of $\boldsymbol{Y}$ would not be much easier. But now we replace $\rho_\beta(z - Z_i)$ with a smooth function $\tilde{R}_i(z)$ such that
$$\tilde{R}_i'(z) \;=\; \begin{cases} z - \beta & \text{if } z \le 0, \\ -\beta & \text{if } 0 \le z \le Z_i - 1, \\ z - Z_i + 1 - \beta & \text{if } Z_i - 1 \le z \le Z_i, \\ 1 - \beta & \text{if } Z_i \le z \le n, \\ z - n + 1 - \beta & \text{if } z \ge n; \end{cases}$$
see also Figure 4. The idea behind $\tilde{R}_i$ is to replace $\rho_\beta(z - Z_i)$ with $\int_{Z_i - 1}^{Z_i} \rho_\beta(z - t)\, dt$, which would result in the derivative $\min\bigl(\max(z - Z_i + 1 - \beta, -\beta), 1 - \beta\bigr)$. The extra modifications on $(-\infty, 0)$ and $(n, \infty)$ are just to ensure the strong coercivity part of Condition (A.3). Thus we propose to minimize
$$\tilde{T}(\boldsymbol{g}) \;:=\; \sum_{i=1}^{n} \tilde{R}_i(g_i) + \sum_{j=1}^{n-1} \lambda_j |g_{j+1} - g_j|$$
by means of Algorithm I and then to utilize the following result.

**Theorem 4.2** *Suppose that $\hat{\boldsymbol{g}}$ minimizes $\tilde{T}$ over $\mathbb{R}^n$. Then $\hat{\boldsymbol{g}} \in (\beta, n - 1 + \beta)^n$. Furthermore, let $\hat{\boldsymbol{f}} \in \mathbb{R}^n$ be given by*
$$\hat{f}_i \;:=\; Y_{(\lceil \hat{g}_i \rceil)}$$
*with the order statistics $Y_{(1)} \le Y_{(2)} \le \cdots \le Y_{(n)}$ of $\boldsymbol{Y}$ and $\lceil a \rceil$ denoting the smallest integer not smaller than $a \in \mathbb{R}$. Then $\hat{\boldsymbol{f}}$ minimizes $T$ over $\mathbb{R}^n$.*

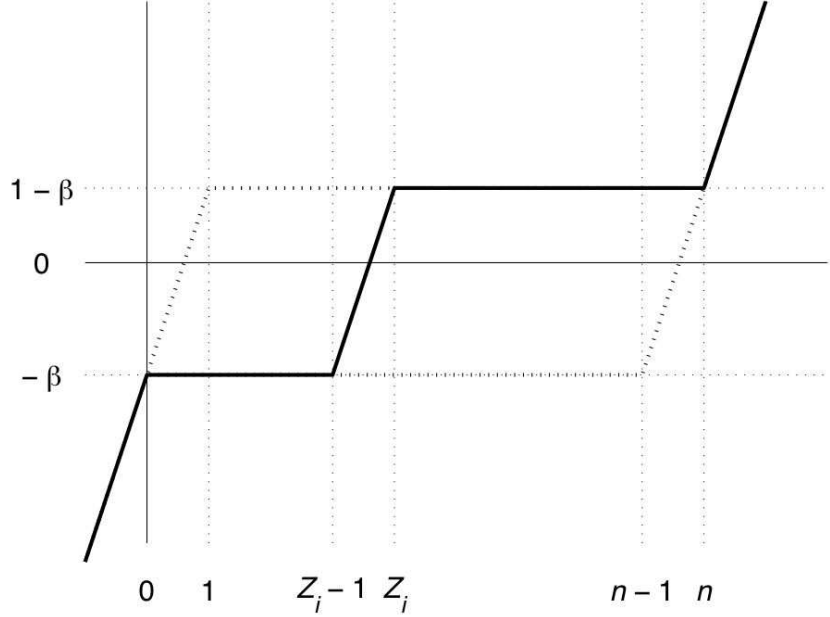

FIG 4. *The derivative $\tilde{R}'_i$.*

**Algorithm II.** Summarizing the preceding findings, we may compute the estimated $\beta$-quantile function by means of Algorithm I, applied to the functions $\tilde{R}'_i$ in place of $R'_i$. Let us just comment on the corresponding auxiliary functions

$$\underline{\tilde{M}}_{jk}(t) := \min\{z \in \mathbb{R} : \tilde{R}'_{jk}(z) \geq t\},$$
$$\overline{\tilde{M}}_{jk}(t) := \max\{z \in \mathbb{R} : \tilde{R}'_{jk}(z) \leq t\}.$$

If $z_{(1)} < z_{(2)} < \cdots < z_{(\ell)}$ are the $\ell := k - j + 1$ order statistics of $(Z_i)_{i=j}^k$, then

$$R'_{jk}(z) = \begin{cases} \ell \min(z, 0) - \ell\beta & \text{if } z \leq z_{(1)} - 1, \\ i + z - z_{(i)} - \ell\beta & \text{if } z_{(i)} - 1 \leq z \leq z_{(i)}, 1 \leq i \leq \ell, \\ i - \ell\beta & \text{if } z_{(i)} \leq z \leq z_{(i+1)} - 1, 1 \leq i < \ell, \\ \ell(1 - \beta + \max(z - n, 0)) & \text{if } z \geq z_{(\ell)}. \end{cases}$$

This entails that

$$\underline{\tilde{M}}_{jk}(t) = \begin{cases} t/\ell + \beta & \text{if } t \leq -\ell\beta, \\ z_{(i)} + t - i + \ell\beta & \text{if } i - 1 - \ell\beta < t \leq i - \ell\beta, 1 \leq i \leq \ell, \\ n + t/\ell - 1 + \beta & \text{if } t > \ell(1 - \beta), \end{cases}$$

$$\overline{\tilde{M}}_{jk}(t) = \begin{cases} t/\ell + \beta & \text{if } t < -\ell\beta, \\ z_{(i)} + t - i + \ell\beta & \text{if } i - 1 - \ell\beta \leq t < i - \ell\beta, 1 \leq i \leq \ell, \\ n + t/\ell - 1 + \beta & \text{if } t \geq \ell(1 - \beta). \end{cases}$$

To implement this algorithm efficiently, one should use two additional vector variables $\mathbf{Z}^{(f)}$ and $\mathbf{Z}^{(g)}$ such that for each segment $\{j,\ldots,k\}$ of $\mathbf{f}$ (resp. $\mathbf{g}$), $(Z_i^{(f)})_{i=j}^k$ (resp. $(Z_i^{(g)})_{i=j}^k$) contains the order statistics of $(Z_i)_{i=j}^k$. Whenever two segments of $\mathbf{f}$ (resp. $\mathbf{g}$) are merged, the vector $\mathbf{Z}^{(f)}$ (resp. $\mathbf{Z}^{(g)}$) may be updated by a suitable version of *MergeSort* (Knuth, 1998).

## 5. The choice of tuning parameters $\lambda_j$

### 5.1. Constant and fixed $\lambda$

Let us first discuss the simple case of a constant value $\lambda > 0$ for all $\lambda_j$. In Example I, let $\hat{\sigma}$ be some consistent estimator of the standard deviation of the variables $Y_i$, assuming temporarily homoscedastic errors $Y_i - \mu(x_i)$. For instance, $\hat{\sigma}$ could be the estimator proposed by Rice (1984) or the version based on the MAD by Donoho et al. (1995). Since $R'_i(z) = z - Y_i$, and since for large $n$ the process

$$[0,1] \ni t \;\mapsto\; n^{-1/2}\hat{\sigma}^{-1/2} \sum_{i \le nt} (\mu(x_i) - Y_i)$$

behaves similarly as a standard Brownian motion by virtue of Donsker's invariance principle, one could use

$$\lambda \;=\; cn^{1/2}\hat{\sigma}$$

for some constant $c > 0$. In our experience with simulated data, a value of $c$ within $[0.15, 0.25]$ yielded often satisfying results.

In Example II, note that the data $Y_i$ may be coupled with independent Bernoulli random variables $\xi_i \in \{0,1\}$ with mean $\beta$ such that

$$\sum_{i=j}^k (\xi_i - \beta) \begin{cases} \ge \sum_{i=j}^k R'_i(Q_\beta(x_i)-) = \sum_{i=j}^k \Big(1\{Y_i < Q_\beta(x_i)\} - \beta\Big) \\ \le \sum_{i=j}^k R'_i(Q_\beta(x_i)+) = \sum_{i=j}^k \Big(1\{Y_i \le Q_\beta(x_i)\} - \beta\Big) \end{cases} \quad (9)$$

for $1 \le j \le k \le n$. Since $t \mapsto n^{-1/2}(\beta(1-\beta))^{-1/2}\sum_{i \le nt}(\xi_i - \beta)$ behaves asymptotically like a standard Brownian motion, too, we propose

$$\lambda \;=\; cn^{1/2}(\beta(1-\beta))^{1/2}.$$

### 5.2. Adaptive choice of the $\lambda_j$

Let $f$ be the unknown underlying regression function. Our goal is to find a "simple" vector (function) $\hat{\mathbf{f}}$ which is adequate for the data in the sense that the deviations between the data and $\hat{\mathbf{f}}$ satisfy a multiresolution criterion (Davies and Kovac, 2001) where we require the deviations between data and $\hat{f}$ at different scales and locations to be no larger than we would expect from noise. More precisely we require for each interval $\{j, \ldots, k\}$ from a collection $\mathcal{I}_n$ of index intervals in $\{1, \ldots, n\}$ that

$$\sum_{i=j}^{k} R'_i(\hat{f}_i+) \ \geq \ \underline{\eta}(\hat{f}_j, \ldots, \hat{f}_k) \quad \text{and} \quad \sum_{i=j}^{k} R'_i(\hat{f}_i-) \ \leq \ \overline{\eta}(\hat{f}_j, \ldots, \hat{f}_k). \tag{10}$$

The bounds $\underline{\eta}(\cdot) < 0 < \overline{\eta}(\cdot)$ to be specified later will be chosen such that the inequalities above are satisfied with high probability in case of replacing $\hat{f}$ with the true regression function $f$. A typical choice for $\mathcal{I}_n$ is the family of all $n(n+1)/2$ such intervals $\{j, \ldots, k\}$. Computational complexity can be reduced by considering a smaller collection such as the family of all intervals with dyadic endpoints,

$$\{2^\ell m + 1, \ldots, 2^\ell(m+1)\},$$

where $0 \leq \ell \leq \lfloor \log_2 n \rfloor$ and $0 \leq m \leq \lfloor 2^{-\ell}(n-1) \rfloor$. The difference in computational speed between these two choices for $\mathcal{I}_n$ is easily noticeable in practice. The effect on the resulting approximation $\hat{f}$, however, is rather small. If a vector $\boldsymbol{g}$ does not approximate the data well on some interval $J$ which is not part of the scheme with the dyadic endpoints, then occasionally the multiscale criterion using the dyadic endpoints will consider $g$ to be adequate where the multiscale criterion which makes use of all subintervals will notice the lack of fit. Since this effect is barely noticeable we prefer to use smaller collections such as the family with dyadic endpoints which was also used in the simulation study in Section 6.

To obtain a vector $\hat{\boldsymbol{f}}$ satisfying (10) for all intervals in $\mathcal{I}_n$, we propose an iterative method for the data-driven choice of the tuning parameters $\lambda_i$. This approach generalizes the local squeezing technique from Davies and Kovac (2001) to our general setting. We start with some constant tuning vector $\boldsymbol{\lambda}^{(1)} = (\Lambda, \Lambda \ldots, \Lambda)$ where $\Lambda$ is chosen so large that the corresponding fit $\hat{\boldsymbol{f}}^{(1)}$ is constant. Now suppose that we have already chosen tuning vectors $\boldsymbol{\lambda}^{(1)}, \ldots, \boldsymbol{\lambda}^{(s)}$, and the corresponding fits are denoted by $\hat{\boldsymbol{f}}^{(1)}, \ldots, \hat{\boldsymbol{f}}^{(s)}$. If $\hat{\boldsymbol{f}}^{(s)}$ is still inadequate for the data, we define $\mathcal{J}^{(s)}$ to be the union of all intervals $\{j-1, j, \ldots, k\}$ such that $\{j, \ldots, k\}$ is an interval in $\mathcal{I}_n$ violating (10) with $\hat{\boldsymbol{f}} = \hat{\boldsymbol{f}}^{(s)}$. Then for some fixed $\gamma \in (0, 1)$, e.g. $\gamma = 0.9$, we define

$$\lambda_i^{(s+1)} \ := \ \begin{cases} \gamma \lambda_i^{(s)} & \text{if } i \in \mathcal{J}^{(s)}, \\ \lambda_i^{(s)} & \text{if } i \notin \mathcal{J}^{(s)}. \end{cases}$$

One can easily derive from (3) that for sufficiently large $s$ the fit $\hat{\boldsymbol{f}} = \hat{\boldsymbol{f}}^{(s)}$ does satisfy (10) for all $\{j, \ldots, k\} \in \mathcal{I}_n$.

**Example I (continued).** In this case the multiresolution criterion (10) inspects the sums of residuals on all intervals $I \in \mathcal{I}_n$. If we assume additive and homoscedastic Gaussian white noise, possible choices for $\underline{\eta}(\cdot)$ and $\overline{\eta}(\cdot)$ are

$$\overline{\eta}(f_j,\ldots,f_k) = -\underline{\eta}(f_j,\ldots,f_k) := \hat{\sigma}\sqrt{k-j+1} \cdot \sqrt{2\log(n)},$$

$$\overline{\eta}(f_j,\ldots,f_k) = -\underline{\eta}(f_j,\ldots,f_k) := \hat{\sigma}\sqrt{k-j+1} \cdot \left(\sqrt{2\log\left(\frac{en}{k-j+1}\right)}+c\right)$$

for some $c \geq 0$. The first proposal coincides exactly with the local squeezing technique by Davies and Kovac (2001). The second one is motivated by results of Dümbgen and Spokoiny (2001).

**Example II (continued).** If we assume that $Y_i = f_i + \varepsilon_i$ where the $\varepsilon_1,\ldots,\varepsilon_n$ are independent with $\beta$–quantile $0$ and continuous distribution function, then both $\sum_{i=j}^{k}(R'_i(f_i+) + \beta)$ and $\sum_{i=j}^{k}(R'_i(f_i-) + \beta)$ are binomially distributed with parameters $k-j+1$ and $\beta$. Let $B(x; N,p)$ be the distribution function of a binomial distribution with parameters $N$ and $p$. Then we define $\overline{\eta}(f_j,\ldots,f_k) = \overline{h}(k-j+1)$ minimal and $\underline{\eta}(f_j,\ldots,f_k) = \underline{h}(k-j+1)$ maximal such that

$$B\big((k-j+1)\beta + \overline{h}(k-j+1); k-j+1,\beta\big) \geq 1-n^{-1} \quad \text{and}$$
$$B\big((k-j+1)\beta + \underline{h}(k-j+1); k-j+1,\beta\big) \leq n^{-1}.$$

**Example III (continued).** We assume that for each $Y_i$ is Poisson distributed with parameter $\exp(f_i)$. Then $\sum_{i=j}^{k} Y_i = \sum_{i=j}^{k}\exp(f_i) - \sum_{i=j}^{k} R'_i(f_i)$ is again Poisson distributed with parameter $\sum_{i=j}^{k}\exp(f_i)$. With $P(\cdot;\ell)$ denoting the distribution function of the Poisson distribution with parameter $\ell$, we define $\overline{\eta}(f_j,\ldots,f_k)$ to be $\overline{h}\big(\sum_{i=j}^{k}\exp(f_i)\big)$ and $\underline{\eta}(f_j,\ldots,j_k) = \underline{h}\big(\sum_{i=j}^{k}\exp(f_i)\big)$, where $\overline{h}(\ell)$ is maximal and $\underline{h}(\ell)$ is minimal such that

$$P(\ell - \overline{h}(\ell); \ell) \geq 1 - n^{-1} \quad \text{and} \quad P(\ell - \underline{h}(\ell); \ell) \leq n^{-1}.$$

**Example IV (continued).** Suppose that $Y_i,\ldots,Y_n$ are binomially distributed with parameters $1$ and $p_i = \exp(f_i)/(1+\exp(f_i))$. Then $\sum_{i=j}^{k} Y_i$ may be written as $\sum_{i=j}^{k} p_i - \sum_{i=j}^{k} R'_i(f_i)$. Following Hoeffding's (1956) finding that the deviations of $\sum_{i=j}^{k} Y_i$ from its mean $\sum_{i=j}^{k} p_i$ tend to be largest in case of equal probabilities $p_i$, we define $\overline{\eta}(f_j,\ldots,f_k) = \overline{h}(k-j+1,\bar{p}_{jk})$ and $\underline{\eta}(f_j,\ldots,j_k) = \underline{h}(k-j+1,\bar{p}_{jk})$, where $\bar{p}_{jk}$ denotes the mean of $p_j,\ldots,p_k$ while $\overline{h}(N,p)$ is maximal and $\underline{h}(N,p)$ is minimal such that

$$B(Np - \overline{h}(N,p); N,p) \geq 1-n^{-1} \quad \text{and} \quad B(Np - \underline{h}(N,p); N,p) \leq n^{-1}.$$

For the consistency results to follow, it is crucial that the adaptive choice of the $\lambda_i$ yields a fit $\hat{f}$ such that for some constant $c_o$,

$$\pm\sum_{i=j}^{k} R'_i(\hat{f}_i \mp) \leq \big(c_o(k-j+1)\log n\big)^{1/2} + c_o\log n \quad \text{for all } 1 \leq j \leq k \leq n. \quad (11)$$

For example I this is obvious, at least if $\mathcal{I}_n$ comprises all subintervals of $\{1,\ldots,n\}$. By means of suitable exponential inequalities one can verify the multiscale criterion (11) for Examples II-IV, too.

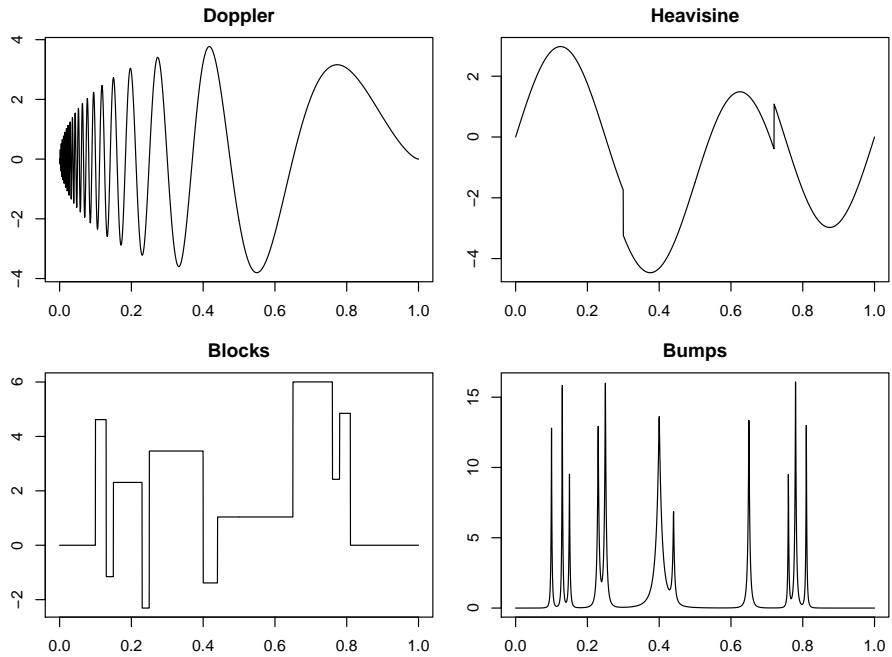

FIG 5. *Rescaled versions of standard test signals by Donoho and Johnstone.*

## 6. Numerical examples

A simulation study was carried out to compare the median number of local extreme values for nine different versions of the general taut string method. Figure 5 shows rescaled versions of four standard test signals by Donoho (1993) and Donoho and Johnstone (1994) that have been used to create samples under four different test beds as described in detail below. For each function $f$ and sample size $n$ the following test beds were considered:

- **Gaussian:** Independent normal observations

$$Y_i \sim \mathcal{N}(f(i/n), 0.4), \quad i = 1, \ldots, n$$

  were generated. The usual taut string method and the quantile version with $\beta = 0.5$ were applied to recover $f$ and the quantile version with $\beta = 0.1$ and $\beta = 0.9$ was used to recover the 0.1– and 0.9–quantile curves of the data which are approximately $f - 0.513$ and $f + 0.513$.

- **Cauchy:** Similarly Cauchy observations were generated by

$$Y_i \sim \mathcal{C}(f(i/n), 0.4), \quad i = 1, \ldots, n,$$

  where $\mathcal{C}(l, s)$ denotes the Cauchy distribution with location $l$ and scale $s$ having density function

$$p(y) = \left(\pi s(1 + ((y-l)/s)^2)\right)^{-1}$$

Since the mean of the Cauchy distribution does not exist, only the quantile taut string was applied with quantiles 0.1, 0.5 and 0.9 to recover $f - 1.231$, $f$ and $f + 1.231$.

- **Binary:** Binary observations were obtained by sampling from a Binomial distribution:

$$Y_i \sim \text{Bin}(1, p_i), \quad p_i = (f(i/n) - a)/(b - a), \quad i = 1, \ldots, n$$

with $b = \max_{t \in [0,1]} f(t)$ and $a = \min_{t \in [0,1]} f(t)$. Then the taut string method for Binary data was used to recover $p_i$.

- **Poisson:** Finally, Poisson data were derived by

$$Y_i \sim \text{Poisson}(l_i), \quad l_i = f(i/n) - a, \quad i = 1, \ldots, n$$

with $a = \min_{t \in [0,1]} f(t)$. Then the taut string method for Poisson data was applied to recover $l_i$.

Typical approximations from samples of size 2048 for the Blocks and the Doppler signals can be seen in Figure 6 and Figure 7. In each Figure the first row illustrates the Gaussian testbed with the usual taut string method in the left and the quantile versions in the right panel. The robust method which corresponds to the 0.5 quantile is plotted in grey, the other quantiles are plotted in black. The Cauchy data are shown in the second row. The left panel demonstrates the huge range of the observations which lie between -137 and 12383. The right panel shows a zoom in and approximation to the quantiles where again the 0.5 quantile is plotted in grey. Finally the last row shows binary and Poisson data.

For each of the four signals, three different sample sizes and each of the four test models 100 samples were generated and the various taut string methods were applied to the data as described above in the description of the test beds. For each application of one of the methods to a sample the number of local extreme values in the approximation was determined. Table 1 reports the median number of local extreme values over the simulations. In brackets the mean absolute deviation from the true number of local extreme values is given apart from the samples derived from the Doppler function which has an infinite number of local extreme values.

These simulations confirm that the usual taut string method is excellent in fitting the correct number of local extreme values and very reliably attains the correct number of local extreme values already for samples of size 512. However, the robust version performs remarkably well in the Gaussian case and has the additional advantage that it depends much less on the distribution of the errors and performs similarly in the Cauchy test bed. In contrast the approximation of 0.1 and 0.9 quantiles is much more difficult. Even for large sample sizes the fitted models often miss local extreme values, in particular for the 0.1 quantile of the Bumps data set which is an extremely difficult situation. The binary problem also appears to be considerably difficult, although still much of the underlying structure is recovered using the 0/1 observations. For the Poisson case the detection rate of the correct number of local extreme values is nearly as good as the robust taut string.

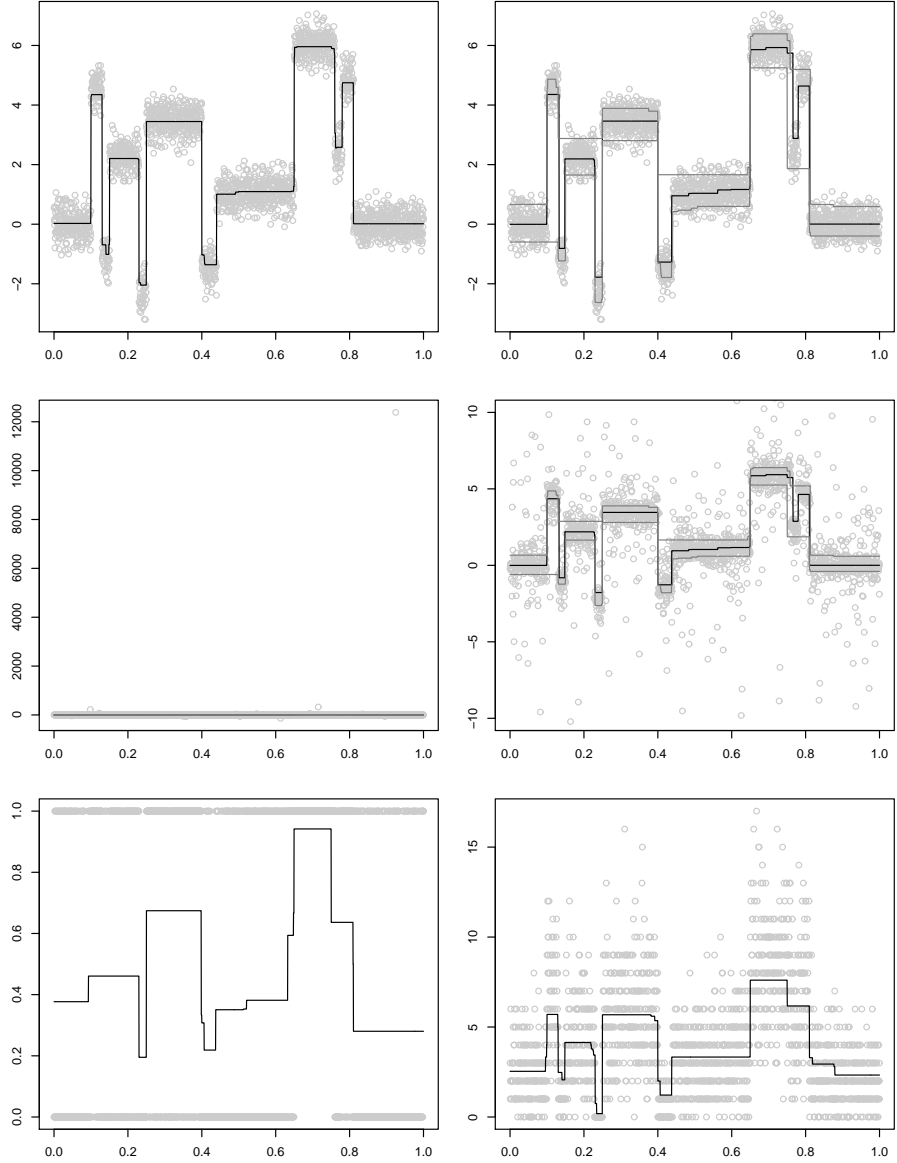

Fig 6. *Typical approximations to the Blocks signal. First row: Approximations to Gaussian data using usual taut string method and quantile versions. Second row: Approximations to Cauchy data, original scale left, zoom in right, Last row: Approximations to binary and Poisson data.*

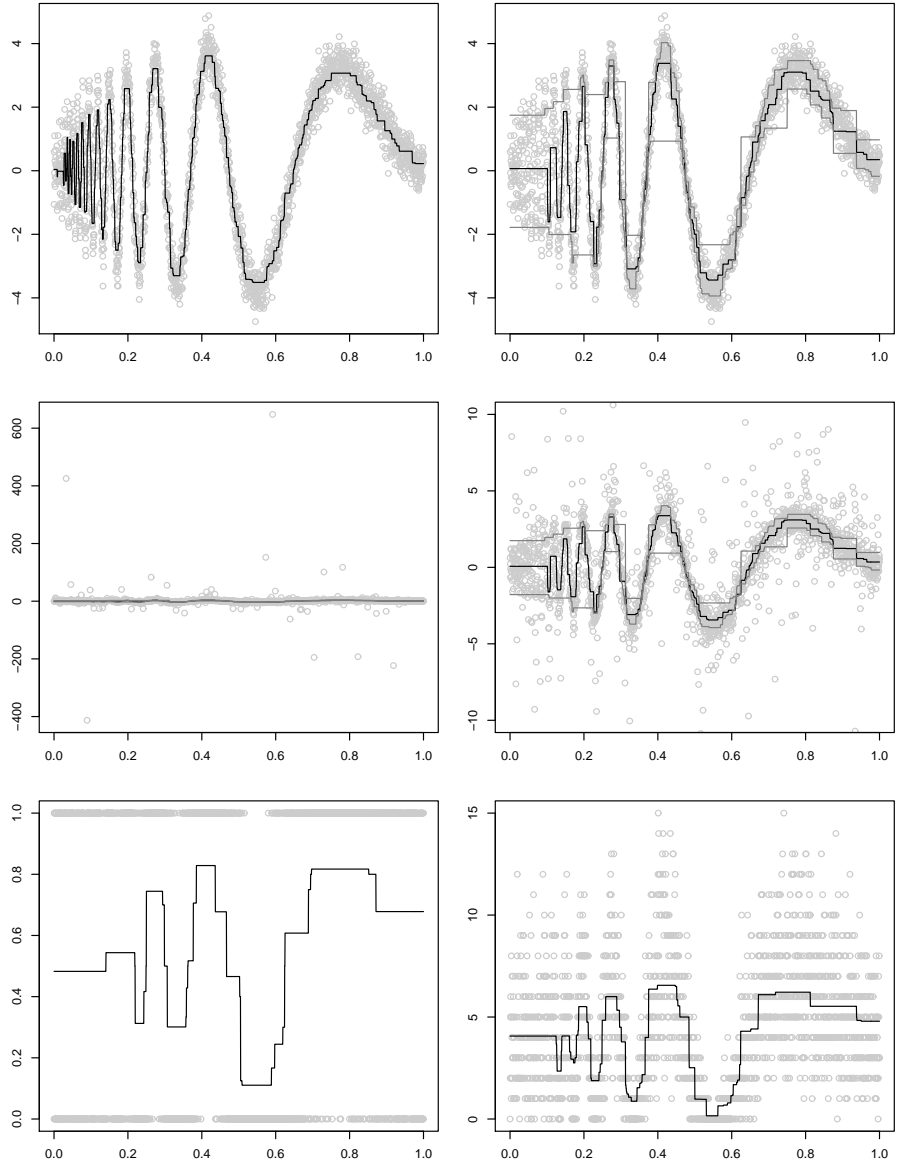

FIG 7. *Typical approximations to the Doppler signal. First row: Approximations to Gaussian data using usual taut string method and quantile versions. Second row: Approximations to Cauchy data, original scale left, zoom in right, Last row: Approximations to binary and Poisson data.*

TABLE 1
*Comparison of the median number of local extreme values for nine different versions of the general taut string method, as described in the text. In brackets the mean absolute deviation from the true number of local extreme values.*

|  |  | Gaussian | | | | Cauchy | | | Binary | Poisson |
|---|---|---|---|---|---|---|---|---|---|---|
| Data | $n$ | usual | robust | 0.1 qnt | 0.9 qnt | robust | 0.1 qnt | 0.9 qnt | | |
| Doppler | 512 | 21 | 6 | 2 | 3 | 4 | 1 | 1 | 3 | 8 |
| (true: ∞) | 2048 | 28 | 12 | 8 | 7 | 10 | 4 | 3 | 7 | 12 |
|  | 8192 | 34 | 19 | 12 | 13 | 19 | 8 | 9 | 11 | 17 |
| Heavisine | 512 | 6 | 4 | 3 | 3 | 4 | 1 | 0 | 2 | 3 |
| (true: 6) |  | (0.6) | (2.0) | (2.9) | (2.9) | (2.4) | (4.5) | (5.3) | (4.0) | (2.6) |
|  | 2048 | 6 | 6 | 4 | 4 | 4 | 3 | 3 | 3 | 4 |
|  |  | (0.0) | (0.8) | (2.0) | (2.0) | (1.8) | (3.3) | (3.2) | (2.7) | (2.0) |
|  | 8192 | 6 | 6 | 6 | 6 | 6 | 3 | 4 | 4 | 4 |
|  |  | (0.0) | (0.0) | (0.9) | (0.0) | (0.0) | (2.5) | (2.5) | (2.0) | (1.9) |
| Blocks | 512 | 9 | 3 | 4 | 3 | 3 | 1 | 0 | 2 | 7 |
| (true: 9) |  | (0.1) | (6.0) | (5.3) | (5.9) | (6.0) | (7.4) | (8.4) | (7.0) | (2.8) |
|  | 2048 | 9 | 9 | 4 | 5 | 9 | 4 | 3 | 5 | 7 |
|  |  | (0.2) | (0.0) | (5.0) | (4.0) | (0.9) | (4.7) | (5.5) | (3.7) | (1.6) |
|  | 8192 | 9 | 9 | 9 | 5 | 9 | 6 | 5 | 9 | 9 |
|  |  | (0.2) | (0.0) | (0.0) | (3.5) | (0.0) | (3.4) | (4.1) | (0.6) | (0.0) |
| Bumps | 512 | 21 | 5 | 0 | 7 | 3 | 0 | 1 | 1 | 13 |
| (true: 21) |  | (0.0) | (16.4) | (21.0) | (15.0) | (18.4) | (21.0) | (19.1) | (19.8) | (6.9) |
|  | 2048 | 21 | 13 | 3 | 11 | 9 | 0 | 9 | 7 | 21 |
|  |  | (0.0) | (8.4) | (18.7) | (9.2) | (11.5) | (20.9) | (11.4) | (13.3) | (0.4) |
|  | 8192 | 21 | 21 | 9 | 21 | 21 | 2 | 19 | 13 | 21 |
|  |  | (0.1) | (0.0) | (11.2) | (0.0) | (0.0) | (18.8) | (2.6) | (7.8) | (0.0) |

## 7. Consistency

In this section we derive consistency results for fitted regression functions $\hat{f}$ on certain intervals on which the fit is monotone while the true regression function $f_*$ satisfies a Hölder condition.

Precisely, we consider a triangular scheme of observations $(x_i, Y_i) = (x_{in}, Y_{in})$ and auxiliary functions $R_i = R_{in}$. Asymptotic statements refer usually to $n \to \infty$, and we use the abbreviation

$$\rho_n := \frac{\log n}{n}.$$

Throughout let $[A, B]$ be a fixed nondegenerate interval such that the following conditions are satisfied:

**(C.1)** There are constants $m_o, m_* > 0$ such that the design measure $M_n := \sum_{i=1}^n \delta_{x_{in}}$ satisfies

$$\frac{M_n[a, b]}{n(b-a)} \geq m_o$$

for sufficiently large $n$ and all $A \leq a < b \leq B$ with $(b - a) \geq m_* \rho_n$.

**(C.2)** For arbitrary indices $i$ and real numbers $t$,

$$\rho'_{in}(t\pm) := \mathbb{E}R'_{in}(t\pm)$$

exist. Moreover, there exists a nondecreasing function $H : [0, \infty) \to [0, \infty)$ with $H(0) = 0$ and $h_o := \liminf_{t \downarrow 0} H(t)/t > 0$ such that for all $s > 0$ and indices $i$ with $x_{in} \in [A, B]$,

$$\pm \rho'_{in}\big((f_*(x_{in}) \pm s)\pm\big) \;\geq\; H(s).$$

**(C.3)** For real numbers $a \leq b$ and $t$ define

$$\Delta_n^{(\pm)}(a, b, t) \;:=\; \sum_{i \,:\, a \leq x_{in} \leq b} (R'_{in} - \rho'_{in})\big((f_*(x_{in}) + t)\pm\big).$$

Then there exist constants $K_1, K_2 > 0$ such that for all $A \leq a \leq b \leq B$ and $\eta \geq 0$,

$$\mathbb{P}\Big(\sup_{t \in \mathbb{R}} \Delta_n^{(\pm)}(a, b, t)^2 > M_n[a, b]\,\eta\Big) \;\leq\; K_1 \exp(-K_2 \eta).$$

Let us comment briefly on these conditions: Condition (C.1) is satisfied if, for instance, all design points $x_{in}$ are contained in $[A, B]$ with $x_{i+1,n} - x_{in} = (B - A)/n$ for $1 \leq i < n$. It also holds true almost surely if $(x_{in})_{i=1}^n$ is the vector of order statistics of $(X_i)_{i=1}^n$, where $X_1, X_2, X_3, \ldots$ are i.i.d. with a Lebesgue density $g$ which is bounded away from zero on $[A, B]$.

As to Conditions (C.2-3), consider first $R_{in}(t) := (t - Y_{in})^2/2$. Then $R'_{in}(t\pm) = t - Y_{in}$ and $\rho'_{in}(t\pm) = t - f_*(x_{in})$. Thus Condition (C.2) is satisfied with $c = 1$, and Condition (C.3) amounts to the errors $\varepsilon_{in} := Y_{in} - \mu(x_{in})$ having subgaussian tails uniformly in $x_{in} \in [A, B]$. In case of $R_{in}(t) := \rho_\beta(t - Y_{in})$, suppose for simplicity that all distribution functions $F(\cdot \,|\, x)$ are continuous. Then $R'_{in}(t+) = 1\{Y_{in} \leq t\} - \beta$, $R'_{in}(t-) = 1\{Y_{in} < t\} - \beta$, while $\rho'_{in}(t\pm) = F(t \,|\, x_{in}) - \beta$. Condition (C.2) is satisfied, for instance, if $Y_{in} = f_*(x_{in}) + \sigma(x_{in})Z_i$ for some bounded function $\sigma(\cdot)$ and i.i.d. random errors $Z_i$ with continuous and strictly positive density. Moreover, it follows from empirical process theory that Condition (C.3) is satisfied with universal constants $K_1$ and $K_2$.

In what follows let $\hat{f}_n$ be any estimator of $f_*$. Our first consistency result applies to isotonic regression estimators as well as taut string estimators with constant tuning vector $\boldsymbol{\lambda}$ via Theorem 2.4. It also applies to the taut string estimators with adaptively chosen tuning vectors $\boldsymbol{\lambda}$ in case of (11).

**Theorem 7.1** *Suppose that Conditions (C.1-3) hold and that $f_*$ is Hölder continuous on $[A, B]$ with exponent $\gamma \in (0, 1]$, i.e.*

$$L \;:=\; \sup_{A \leq x < y \leq B} \frac{|f_*(y) - f_*(x)|}{(y - x)^\gamma} \;<\; \infty.$$

*Let $[A_n, B_n]$ be a fixed or random subinterval of $[A, B]$ with $B_n - A_n \geq m_* \rho_n$ such that $\hat{f}_n$ is isotonic on $[A_n, B_n]$. Further suppose that either $\hat{f}$ minimizes the sum $\sum_{i \,:\, x_{in} \in [A_n, B_n]} R_{in}(f(x_{in}))$ over all isotonic functions $f : [A_n, B_n] \to \mathbb{R}$ or that $\hat{\boldsymbol{f}}_n = \big(\hat{f}_n(x_{in})\big)_{i=1}^n$ satisfies (11) for all $n$. Then for $\delta_n := \rho_n^{1/(2\gamma+1)}$ and*

*sufficiently large C,*

$$\left.\begin{array}{c}\sup_{x\in[A_n,B_n-\delta_n]}\bigl(\hat{f}(x)-f_*(x)\bigr)^+\\ \sup_{x\in[A_n+\delta_n,B_n]}\bigl(f_*(x)-\hat{f}(x)\bigr)^+\end{array}\right\} \leq C\rho_n^{\gamma/(2\gamma+1)}$$

*with asymptotic probability one.*

Our second consistency result concerns estimation of $f_*$ close to its local extrema. It is known that the least squares taut string estimators tend to underestimate $f_*$ near local maxima and overestimate $f_*$ near local minima. The generalized estimators discussed here have the same property, but this effect can be bounded:

**Theorem 7.2** *Suppose that Conditions (C.1-3) hold. Let $x_* \in (A, B)$ be a local extremum of $f_*$ such that*

$$\limsup_{x \to x_*} \frac{|f_*(x_*) - f_*(x)|}{|x - x_*|^\kappa} < \infty \tag{12}$$

*for some $\kappa > 0$.*

*(a) If $\hat{f}_n$ is a taut string estimator with tuning constants $\lambda_{in}$ in $(0, c_o n^{1/2}]$ for some constant $c_o$, then*

$$\max_{x \in [x_* \pm n^{-1/(2\kappa+2)}]} \hat{f}_n(x) \geq f(x_*) + O_p\bigl(n^{-\kappa/(2\kappa+2)}\bigr),$$
$$\min_{x \in [x_* \pm n^{-1/(2\kappa+2)}]} \hat{f}_n(x) \leq f(x_*) + O_p\bigl(n^{-\kappa/(2\kappa+2)}\bigr).$$

*(b) Let $\hat{f}_n$ be any estimator such that $\hat{\boldsymbol{f}}_n = \bigl(\hat{f}_n(x_{in})\bigr)_{i=1}^n$ satisfies (11) for all n. Then for C sufficiently large,*

$$\max_{x \in [x_* \pm n^{-1/(2\kappa+1)}]} \hat{f}_n(x) \geq f(x_*) - C\rho_n^{\kappa/(2\kappa+1)},$$
$$\min_{x \in [x_* \pm n^{-1/(2\kappa+1)}]} \hat{f}_n(x) \leq f(x_*) + C\rho_n^{\kappa/(2\kappa+1)}$$

*with asymptotic probability one.*

To illustrate the latter results, suppose that $f_*$ is twice differentiable with bounded second derivative on $[A, B]$. If $x_* \in (A, B)$ is a local extremum of $f_*$, then (12) holds true with $\kappa = 2$. Then the taut string estimator $\hat{f}_n$ with global tuning parameter $\lambda = \lambda_n = O(n^{1/2})$ underestimates (resp. overestimates) a local maximum (resp. minimum) by $O_p(n^{-1/3})$. In case of the adaptively chosen $\lambda_{in} = \lambda_{in}(\text{data})$, the latter rate improves to $O_p(\rho_n^{2/5})$.

## 8. Proofs

**Proof of Lemma 2.1.** As mentioned earlier, the necessity of Condition (3) follows from (2) applied to $\pm\boldsymbol{\delta}^{(jk)}$. On the other hand, it will be shown below that an arbitrary vector $\boldsymbol{\delta} \in \mathbb{R}^n$ may be written as

$$\boldsymbol{\delta} \ = \ \sum_{1 \le j \le k \le n} \alpha^{(jk)} \boldsymbol{\delta}^{(jk)}$$

with real numbers $\alpha^{(jk)}$ satisfying the following two constraints:
(i) For $1 \le i \le n$ it follows from $\delta_i > 0$ (or $\delta_i < 0$) that $\alpha^{(jk)} \ge 0$ (or $\alpha^{(jk)} \le 0$) whenever $i \in \{j, \ldots, k\}$.
(ii) For $1 \le j < n$,

$$|\delta_{j+1} - \delta_j| \ = \ \sum_{k=1}^{j} |\alpha^{(kj)}| + \sum_{k=j+1}^{n} |\alpha^{(jk)}|.$$

With this particular representation of $\boldsymbol{\delta}$, one can easily show that

$$DT(\hat{\boldsymbol{f}}, \boldsymbol{\delta}) \ = \ \sum_{1 \le j \le k \le n} |\alpha^{(jk)}| DT(\hat{\boldsymbol{f}}, \operatorname{sign}(\alpha^{(jk)}) \boldsymbol{\delta}^{(jk)}) \ \ge \ 0.$$

The coefficients $\alpha^{(jk)}$ may be constructed iteratively as follows: Let $\mathcal{J}_0 := \{i : \delta_i > 0\}$ and $a_0 := \min\{\delta_i : i \in \mathcal{J}_0\}$. For any maximal index interval $\{j, \ldots, k\} \subset \mathcal{J}_0$ set $\alpha^{(jk)} := a_0$. Then define $\mathcal{J}_1 := \{i : \delta_i > a_0\}$ and $a_1 := \min\{\delta_i - a_0 : i \in \mathcal{J}_1\}$. For any maximal index interval $\{j, \ldots, k\} \subset \mathcal{J}_1$ set $\alpha^{(jk)} := a_1$. Then define $\mathcal{J}_2 := \{i : \delta_i > a_1\}$ and proceed analogously, until we end up with an empty set $\mathcal{J}_\ell$. Similarly, one may start with $\mathcal{K}_0 := \{i : \delta_i < 0\}$, $b_0 := \max\{\delta_i : i \in \mathcal{K}_0\}$, and define $\alpha^{(jk)}$ for selected index intervals $\{j, \ldots, k\}$. □

**Proof of Lemma 2.2.** The necessity of Condition (4) follows from (2) if applied to $\pm \boldsymbol{\delta}^{(1k)}$.

It remains to be shown that any vector $\hat{\boldsymbol{f}}$ satisfying (4) for $1 \le k \le n$ satisfies (2) as well. Note that for any $\boldsymbol{\delta} \in \mathbb{R}^n$,

$$\begin{aligned}
\sum_{i=1}^{n} R'_i(\hat{f}_i) \delta_i &= -\sum_{i=1}^{n} (\delta_n - \delta_i) R'_i(\hat{f}_i) \\
&= -\sum_{i=1}^{n-1} \sum_{k=i}^{n-1} (\delta_{k+1} - \delta_k) R'_i(\hat{f}_i) \\
&= -\sum_{k=1}^{n-1} (\delta_{k+1} - \delta_k) \sum_{i=1}^{k} R'_i(\hat{f}_i),
\end{aligned}$$

since $\sum_{i=1}^{n} R'_i(\hat{f}_i) = 0$. Consequently,

$$DT(\hat{\boldsymbol{f}}, \boldsymbol{\delta}) \ = \ \sum_{k=1}^{n-1} |\delta_{k+1} - \delta_k| H_k,$$

where

$$H_k \ := \ \operatorname{sign}(\delta_{k+1} - \delta_k) \left( \lambda_k \operatorname{sign}(\hat{f}_{k+1} - \hat{f}_k) - \sum_{i=1}^{k} R'_i(\hat{f}_i) \right) + \lambda_k 1\{\hat{f}_{k+1} = \hat{f}_k\}.$$

But condition (4) entails that all these quantities $H_k$ are nonnegative. □

**Proof of Lemma 2.3.** Let $\mathcal{J} = \{j, \ldots, k\}$ a maximal index interval such that $\hat{f}_j = \cdots = \hat{f}_k = \max_i \hat{f}_i$. Then it follows from Lemma 2.2 that

$$\sum_{i=j}^{k} R'_i(\hat{f}_j) = -(\lambda_{j-1} + \lambda_k).$$

If $j > 1$ or $k < n$, the right hand side is strictly negative, whence $\hat{f}_j < z_r$. If $j = 1$ and $k = n$, then $\hat{f}_1 < z_r$, because $\sum_{i=1}^{n} R'_i(z_r) > 0$. These considerations show that $\max_i \hat{f}_i < z_r$, and analogous arguments reveal that $\min_i \hat{f}_i > z_\ell$. □

Our proof of Theorem 2.4 relies on a characterization of isotonic fits which is of independent interest.

**Theorem 8.1** Let $1 \le a \le b \le n$. A vector $(\check{f}_i)_{i=a}^{b}$ minimizes $\sum_{i=a}^{b} R_i(\check{f}_i)$ under the constraint that $\check{f}_a \le \cdots \le \check{f}_b$ if, and only if, for arbitrary $a \le j \le k \le b$,

$$\sum_{i=j}^{k} R'_i(\check{f}_i -) \le 0 \quad \text{whenever } \check{f}_{j-1} < \check{f}_j = \check{f}_k, \tag{13}$$

$$\sum_{i=j}^{k} R'_i(\check{f}_i +) \ge 0 \quad \text{whenever } \check{f}_j = \check{f}_k < \check{f}_{k+1}. \tag{14}$$

Here $\check{f}_{a-1} := -\infty$ and $\check{f}_{b+1} := \infty$.

**Proof of Theorem 8.1.** For notational convenience let $a = 1$ and $b = n$. Note that the functional $f \mapsto T_\uparrow(f) := \sum_{i=1}^{n} R_i(f_i)$ is convex on $\mathbb{R}^n$, and that the set $\mathbb{R}^n_\uparrow$ of vectors in $\mathbb{R}^n$ with non-decreasing components is convex. Thus an isotonic vector $\check{f}$ minimizes $T_\uparrow$ over $\mathbb{R}^n_\uparrow$ if, and only if,

$$DT_\uparrow(\check{f}, \delta) = \sum_{i=1}^{n} R'_i(\check{f}_i, \operatorname{sign}(\delta_i))\delta_i \ge 0$$

for any $\delta \in \mathbb{R}^n$ such that $\check{f} + t\delta$ is isotonic for some $t > 0$. The latter requirement is equivalent to

$$\delta_i \le \delta_j \quad \text{whenever } i < j \text{ and } \check{f}_i = \check{f}_j. \tag{15}$$

Condition (15) is satisfied for $\delta = -\delta^{(jk)}$ if $\check{f}_{j-1} < \check{f}_j$, and for $\delta = \delta^{(jk)}$ if $\check{f}_k > \check{f}_{k+1}$. Thus the conditions stated in Theorem 8.1 are necessary.

On the other hand, on can easily show that any $\delta$ satisfying (15) may be written as a sum $\sum_{1 \le j \le k \le n} \alpha^{(jk)} \delta^{(jk)}$ with real numbers $\alpha^{(jk)}$ satisfying (i) in the proof of Lemma 2.1 and

(iii) $\alpha^{(jk)} < 0$ (resp. $\alpha^{(jk)} > 0$) implies that $\check{f}_{j-1} < \check{f}_j$ (resp. $\check{f}_k < \check{f}_{k+1}$).

One can deduce from this representation that

$$DT_\uparrow(\check{f}, \delta) = \sum_{1 \le j \le k \le n} |\alpha^{(jk)}| DT_\uparrow(\check{f}, \operatorname{sign}(\alpha^{(jk)})\delta^{(jk)}),$$

and each summand on the right hand side is non-negative by (13) and (14). □

**Proof of Theorem 2.4.** We have to verify Conditions (13) and (14) with $\check{f}_i = \hat{f}_i$ for $a \leq i \leq b$. But it follows from (3) and our assumptions on $\boldsymbol{\lambda}$ that the sum in (13) is not greater than $\lambda_{j-1}\overline{\text{sign}}(\hat{f}_{j-1} - \hat{f}_j) + \lambda_k\overline{\text{sign}}(\hat{f}_{k+1} - \hat{f}_k) = 0$, while the sum in (14) is not smaller than $\lambda_{j-1}\underline{\text{sign}}(\hat{f}_{j-1} - \hat{f}_j) + \lambda_k\underline{\text{sign}}(\hat{f}_{k+1} - \hat{f}_k) = 0$. □

**Proof of Lemma 3.1.** Suppose first that $c \geq \overline{M}_{\ell m}(v)$. For any $z > c$,

$$R'_{jm}(z) = \underbrace{R'_{jk}(z)}_{\geq u} + \underbrace{R'_{\ell m}(z)}_{> v} > u + v,$$

whence $\overline{M}_{jm}(u+v) \leq c$. Moreover,

$$R'_{jm}(\overline{M}_{\ell m}(v)) = \underbrace{R'_{jk}(\overline{M}_{\ell m}(v))}_{\leq u} + \underbrace{R'_{\ell m}(\overline{M}_{\ell m}(v))}_{=v} \leq u + v,$$

so that $\overline{M}_{jm}(u+v) \geq \overline{M}_{\ell m}(v)$. This proves Part (a).

As for Part (b), suppose that $c > \overline{M}_{jm}(u+v)$. Then for $c \geq z > \overline{M}_{jm}(u+v)$,

$$u + v < R'_{jm}(z) = R'_{jk}(z) + R'_{\ell m}(z) \leq u + R'_{\ell m}(z),$$

so that $R'_{\ell m}(z) > v$. Hence $\overline{M}_{\ell m}(v) \leq \overline{M}_{jm}(u+v)$. □

**Proof of Theorem 3.2.** Let $\mathcal{J} = \{j, \ldots, k\}$ be a local maximum of $\hat{\boldsymbol{f}}$. A close inspection of Algorithm I reveals that

$$\hat{f}_i = \underline{M}_{jk}(-\lambda_{j-1} - \lambda_k) \quad \text{for } i \in \mathcal{J}.$$

On the other hand, it follows from our assumption on $\boldsymbol{f}$ that

$$R'_{jk}\left(\max_{i \in \mathcal{J}} f_i\right) \geq \sum_{i=j}^{k} R'_i(f_i) \geq -\lambda_{j-1} - \lambda_k,$$

so that

$$\max_{i \in \mathcal{J}} f_i \geq \max_{i \in \mathcal{J}} \hat{f}_i.$$

Analogously one can show that $\min_{i \in \mathcal{K}} f_i \leq \min_{i \in \mathcal{K}} \hat{f}_i$ for any local minimum $\mathcal{K}$ of $\hat{\boldsymbol{f}}$. □

**Proof of Theorem 4.1.** It follows from Assumption (7) that $T_\varepsilon$ converges pointwise to $T$ as $\varepsilon \downarrow 0$. Since all functions $T$ and $T_\varepsilon$ are convex, it is well-known from convex analysis that the convergence is even uniform on arbitrary compact sets. Specifically consider the closed ball $B_R(\boldsymbol{0})$ around $\boldsymbol{0}$ with radius $R > 0$. It follows from (A.2) that for suitable $R > 0$,

$$T(\boldsymbol{0}) < \min_{\boldsymbol{f} \in \partial B_R(\boldsymbol{0})} T(\boldsymbol{f}).$$

Hence for some $\varepsilon_o > 0$,

$$T_\varepsilon(\mathbf{0}) < \min_{\mathbf{f} \in \partial B_R(\mathbf{0})} T_\varepsilon(\mathbf{f}) \quad \text{for } 0 < \varepsilon \leq \varepsilon_o.$$

These inequalities and convexity of the functions $T$ and $T_\varepsilon$ together entail that the sets $\hat{\mathcal{F}}$ and $\hat{\mathcal{F}}_\varepsilon$, $0 < \varepsilon \leq \varepsilon_o$, are nonvoid and compact subsets of $B_R(\mathbf{0})$. Now convergence of $\hat{\mathcal{F}}_\varepsilon$ to $\hat{\mathcal{F}}$ follows easily from

$$\hat{\mathcal{F}}_\varepsilon \subset \left\{ \mathbf{f} \in B_R(\mathbf{0}) : T(\mathbf{f}) \leq \min_{\mathbf{g} \in B_R(\mathbf{0})} T(\mathbf{g}) + 2 \max_{\mathbf{g} \in B_R(\mathbf{0})} |T(\mathbf{g}) - T_\varepsilon(\mathbf{g})| \right\}. \quad \Box$$

**Proof of Theorem 4.2.** That $\hat{\mathbf{g}} \in (\beta, n-1+\beta)^n$ follows from Lemma 2.3 and the fact that $\tilde{R}'_i(\beta) \leq 0$ with strict inequality if $Z_i > 1$, while $\tilde{R}'_i(n-1+\beta) \geq 0$ with strict inequality if $Z_i < n$. Thus $\hat{\mathbf{f}}$ is well-defined, and it suffices to verify (3) for indices $1 \leq j \leq k \leq n$. But the definitions of $\hat{\mathbf{f}}$ and $\tilde{R}'_i$ entail that

$$\sum_{i=j}^k R'_i(\hat{f}_i +) = \sum_{i=j}^k (1\{Y_i \leq \hat{f}_i\} - \beta)$$
$$\geq \sum_{i=j}^k (1\{Z_i \leq \lceil \hat{g}_i \rceil\} - \beta)$$
$$\geq \sum_{i=j}^k \tilde{R}'_i(\hat{g}_i).$$

According to Lemma 2.1, applied to $\tilde{T}$ in place of $T$, the right hand side is not smaller than

$$\lambda_{j-1}\underline{\text{sign}}(\hat{g}_{j-1} - \hat{g}_j) + \lambda_k\underline{\text{sign}}(\hat{g}_{k+1} - \hat{g}_k)$$
$$= \lambda_{j-1}(1 - 2 \cdot 1\{\hat{g}_{j-1} \leq \hat{g}_j\}) + \lambda_k(1 - 2 \cdot 1\{\hat{g}_{k+1} \leq \hat{g}_k\})$$
$$\geq \lambda_{j-1}(1 - 2 \cdot 1\{\hat{f}_{j-1} \leq \hat{f}_j\}) + \lambda_k(1 - 2 \cdot 1\{\hat{f}_{k+1} \leq \hat{g}_k\})$$
$$= \lambda_{j-1}\underline{\text{sign}}(\hat{f}_{j-1} - \hat{f}_j) + \lambda_k\underline{\text{sign}}(\hat{f}_{k+1} - \hat{f}_k).$$

This proves the first part of (3). Similarly,

$$\sum_{i=j}^k R'_i(\hat{f}_i -) = \sum_{i=j}^k (1\{Y_i < \hat{f}_i\} - \beta)$$
$$\leq \sum_{i=j}^k (1\{Z_i < \lceil \hat{g}_i \rceil\} - \beta)$$
$$\leq \sum_{i=j}^k \tilde{R}'_i(\hat{g}_i)$$
$$\leq \lambda_{j-1}\overline{\text{sign}}(\hat{g}_{j-1} - \hat{g}_j) + \lambda_k\overline{\text{sign}}(\hat{g}_{k+1} - \hat{g}_k)$$
$$\leq \lambda_{j-1}\overline{\text{sign}}(\hat{f}_{j-1} - \hat{f}_j) + \lambda_k\overline{\text{sign}}(\hat{f}_{k+1} - \hat{f}_k). \quad \Box$$

**Proof of Theorem 7.1.** It follows from condition (C.3) that

$$\sup_{A \leq a \leq b \leq B, t \in \mathbb{R}} \Delta_n^{(\pm)}(a, b, t)^2 \leq \eta_o M_n[a, b] \log n \tag{16}$$

with probability at least

$$1 - n^2 K_1 \exp(-K_2 \eta_o \log n) \to 1 \quad \text{if } \eta_o > 2/K_2.$$

Suppose that $\hat{f}_n(x_n) > f_*(x_n) + \varepsilon_n$ for some $x_n \in [A_n, B_n - \delta_n]$ and $\varepsilon_n = C\rho_n^{\gamma/(2\gamma+1)} = C\delta_n^\gamma$ with $C$ to be specified later. Then for $x_n \leq x \leq x_n + \delta_n$,

$$\hat{f}_n(x) - f_*(x) \geq \hat{f}_n(x_n) - f_*(x_n) - |f_*(x) - f_*(x_n)| > (C - L)\delta_n^\gamma.$$

If $\hat{f}_n$ minimizes the sum $\sum_{i : A_n \leq x_{in} \leq B_n} R_i(f(x_{in}))$ over all isotonic functions $f$ on $[A_n, B_n]$, we assume without loss of generality that $\hat{f}_n(x_{in}) < \hat{f}(x_n)$ whenever $A_n \leq x_{in} < x_n$. For otherwise we could replace $x_n$ with the smallest design point $x_{in}$ in $[A_n, B_n]$ such that $\hat{f}_n(x_{in}) = \hat{f}_n(x_n)$. Then Theorem 8.1 entails that

$$
\begin{aligned}
0 &\geq \sum_{i : x_n \leq x_{in} \leq x_n + \delta_n} R'_{in}(\hat{f}_n(x_{in})-) \\
&\geq \sum_{i : x_n \leq x_{in} \leq x_n + \delta_n} R'_{in}\big((f_*(x_{in}) + (C - L)\delta_n^\gamma)+\big) \\
&= \sum_{i : x_n \leq x_{in} \leq x_n + \delta_n} \rho'_{in}\big((f_*(x_{in}) + (C - L)\delta_n^\gamma)+\big) \\
&\qquad - \big(\eta_o M_n[x_n, x_n + \delta_n] \log n\big)^{1/2} \\
&\geq H((C - L)\delta_n^\gamma) M_n[x_n, x_n + \delta_n] - \big(\eta_o M_n[x_n, x_n + \delta_n] \log n\big)^{1/2}
\end{aligned}
$$

in case of $C > L$ and (16). If $\hat{f}_n$ is an arbitrary estimator satisfying (11), then only

$$
\begin{aligned}
&\big(c_o M_n[x_n, x_n + \delta_n] \log n\big)^{1/2} + c_o \log n \\
&\geq \sum_{i : x_n \leq x_{in} \leq x_n + \delta_n} R'_{in}(\hat{f}_n(x_{in})-) \\
&\geq H((C - L)\delta_n^\gamma) M_n[x_n, x_n + \delta_n] - \big(\eta_o M_n[x_n, x_n + \delta_n] \log n\big)^{1/2}.
\end{aligned}
$$

But $\delta_n \geq m_* \rho_n$ for sufficiently large $n$, and then the preceding displayed inequalities entail that

$$
\begin{aligned}
H((C - L)\delta_n^\gamma) &\leq \big((\eta_o/m_o)^{1/2} + (c_o/m_o)^{1/2}\big)(\rho_n/\delta_n)^{1/2} + (c_o/m_o)(\rho_n/\delta_n) \\
&= \big((\eta_o/m_o)^{1/2} + (c_o/m_o)^{1/2} + o(1)\big)\delta_n^\gamma.
\end{aligned}
$$

Hence $C \leq L + \big((\eta_o/m_o)^{1/2} + (c_o/m_o)^{1/2} + o(1)\big)/h_o$.

The assertion about the maximum of $f_* - \hat{f}_n$ on the interval $[A_n + \delta_n, B_n]$ is proved analogously. □

**Proof of Theorem 7.2.** We only prove the assertion about the minimum of $\hat{f}_n$ over a neighborhood of $x_*$, because the other part follows analogously. Suppose that $\hat{f}_n > f_* + \varepsilon_n$ on $[x_* \pm \delta_n]$, where both $\delta_n > 0$ and $\varepsilon_n > 0$ are fixed numbers tending to zero, while $\delta_n \geq m_* \rho_n$.

In case of a taut string estimator with parameters $\lambda_{in} \in (0, c_o n^{-1/2}]$, it follows from (3) and (16) that

$$
\begin{aligned}
2c_o n^{1/2} &\geq \sum_{i\,:\,|x_{in}-x_*|\leq \delta_n} R_i'(\hat{f}_n(x_{in})-) \\
&\geq \sum_{i\,:\,|x_{in}-x_*|\leq \delta_n} R_i'\big((f_*(x_{in}) + \varepsilon_n)+\big) \\
&\geq M_n[x_* \pm \delta_n] H(\varepsilon_n) + \Delta_n(x_* - \delta_n, x_* + \delta_n, \varepsilon_n) \\
&\geq M_n[x_* \pm \delta_n] H(\varepsilon_n) - O_p\big(M_n[x_* \pm \delta_n]^{1/2}\big) \\
&\geq (2h_o + o(1)) m_o n \delta_n \varepsilon_n - O_p(n^{1/2}),
\end{aligned}
$$

where $h_o := \liminf_{t\downarrow 0} H(t)/t$. Hence

$$\varepsilon_n \leq O_p\big(n^{-1/2}\delta_n^{-1}\big).$$

On the other hand,

$$\sup_{x\in[x_*\pm\delta_n]} |f_*(x) - f_*(x_*)| \leq O(\delta_n^\kappa). \tag{17}$$

Hence setting $\delta_n := n^{-1/(2\kappa+2)}$ yields the assertion.

In case of any estimator satisfying (11),

$$
\begin{aligned}
\big(c_o M_n[x_* \pm \delta_n] \log n\big)^{1/2} &+ c_o \log n \\
&\geq \sum_{i\,:\,|x_{in}-x_*|\leq \delta_n} R_i'(\hat{f}_n(x_{in})-) \\
&\geq M_n[x_* \pm \delta_n] H(\varepsilon_n) - \big(\eta_o M_n[x_* \pm \delta_n]^{1/2} \log n\big),
\end{aligned}
$$

i.e. $\varepsilon_n$ equals

$$O\Big(\big(\log(n)/M_n[x_* \pm \delta_n]\big)^{1/2} + \log(n)/M_n[x_* \pm \delta_n]\Big) = O\big((\rho_n/\delta_n)^{1/2} + \rho_n/\delta_n\big).$$

Comparing this with (17) shows that one should take $\delta_n = \rho_n^{1/(2\kappa+1)}$, and this yields the assertion about the minimum of $\hat{f}_n$. □

**Software**

The generalized taut string algorithm has been implemented in the ftnonpar package for the statistics software R (Ihaka and Gentleman, 1996). This add-on package can be downloaded and installed by the standard `install.packages()` command of R. All examples considered in this paper are available via the general `genpmreg` function using the `method` parameter to choose from the usual taut string method, the quantile version and the versions for binomial and Poisson noise.